\documentclass[useAMS,usenatbib]{mnras}

\usepackage{graphicx}   
\usepackage{amsmath}    
\usepackage{amssymb}    
\usepackage{multicol}   
\usepackage{bm}         
\usepackage{pdflscape}  
\usepackage{float}      
\usepackage{tipa}       
\usepackage{color, soul}
\soulregister\cite7
\soulregister\ref7
\soulregister\pageref7

\hypersetup{final}

\title[]
{3D MHD simulations and synthetic radio emission from an oblique rotating magnetic massive star}
\author[]{Daley-Yates S., Stevens I. R., A. ud-Doula}

\author[S. Daley-Yates, I. R. Stevens, A. ud-Doula]{S. Daley-Yates$^{1, 2}$\thanks{E-mail:
simon.daley@cea.fr}, I. R. Stevens$^{2}$, A. ud-Doula$^{3}$\\
	$^{1}$Maison de la Simulation, CEA, CNRS, Univ. Paris-Sud, UVSQ, Université Paris-Saclay, F-91191 Gif-sur-Yvette, France\\
$^{2}$School of Physics and Astronomy, University of Birmingham, Edgbaston, Birmingham, B15 2TT, UK\\
$^{3}$Penn State Scranton, 120 Ridge View Drive, Dunmore, PA 18512, USA\\}
\begin{document}

\date{}

\pagerange{\pageref{firstpage}--\pageref{lastpage}} \pubyear{2018}

\maketitle

\label{firstpage}

\begin{abstract}
We have performed 3D isothermal MHD simulation of a magnetic rotating massive star with a non-zero dipole obliquity and predicted the radio/sub-mm observable lightcurves and continuum spectra for a frequency range compatible with ALMA. From these results we also compare the model input mass-loss to that calculated from the synthetic thermal emission. Spherical and cylindrical symmetry is broken due to the obliquity of the stellar magnetic dipole resulting in an inclination and phase dependence of both the spectral flux and inferred mass-loss rate, providing testable predictions of variability for oblique rotator. Both quantities vary by factors between 2 and 3 over a full rotational period of the star, demonstrating that the role of rotation as critical in understanding the emission. This illustrates the divergence from a symmetric wind, resulting in a two armed spiral structure indicative of a oblique magnetic rotator. We show that a constant spectral index, $\alpha$, model agrees well with our numerical prediction for a spherical wind for $\nu~<~10^{3} \ \mathrm{GHz}$, however it is unable to capture the behavior of emission at $\nu~>~10^{3} \ \mathrm{GHz}$. As such we caution the use of such constant $\alpha$ models for predicting emission from non-spherical winds such as those which form around magnetic massive stars.
\end{abstract}


\section{Introduction} 
\label{sec:OstarMHD_intro}

Magnetism in massive stars has gained significant attention in recent decades due to the unexpected number of these stars that display global, dynamically significant, magnetic fields \citep{Petit2013, Wade2013, Wade2015, Wade2016}. They are unexpected since massive stars have their convective zones beneath a radiative outer envelope; inhibiting the dynamo action thought responsible for generating stellar magnetic fields (see \cite{Walder2012} for a recent review). \cite{Cantiello2009a} report theoretical results which indicate the coupling of subsurface convection with wind clumping and emergence of magnetic field on the surface of O and B-type stars. This picture has proved too simplistic for explaining the approximately 10\% of Galactic O-type and B-type stars with detectable magnetic fields \citep{Wade2013}. The MiMeS project (Magnetism in Massive Stars) has done considerable work advancing our understanding of these stars both from an observational and theoretical stand point \citep{Wade2016}. Now the BinaMIcs project (Magnetism in Massive Stars and Binarity and Magnetic Interactions) is illustrating that massive star magnetism occurs in binary systems as well \citep{Alecian2015}, with $\varepsilon$ Lupi the first discovered magnetic massive binary where both the primary and secondary possess detectable magnetic fields \citep{Shultz2015}. Theoretical and numerical studies are required to help understand the wealth of observational data coming from these projects.

Analytic and semi-analytic modelling of the inner magnetospheres of massive stars have been conducted by \cite{Townsend2005} and \cite{Townsend2007} producing the Rigidly Rotating Magnetosphere model (RRM) and \cite{Owocki2016} producing the Analytic Dynamical Magnetosphere model (ADM). Both models are designed to overcome the limitations of direct numerical simulations and to provide insight without the computationally intensive treatment of full MHD. These analytic models capture the suspension of material on magnetospheric field lines and while successful at reproducing observable emission in X-rays and H$_{\alpha}$, they are unable to describe the free streaming wind or magnetically perturbed material at large radii. As such they are unsuitable for predicting radio/sub-mm emission.

\cite{ud-Doula2002} first studied the role of large scale cylindrically symmetric magnetic fields in shaping the dynamics and structure of massive star winds using 2D isothermal Magnetohydrodynamic (MHD) simulations. Results showed coherent disk structures forming in the magnetic equator as outflowing wind material is channelled by the magnetic field lines forming a standing shock. This model was improved upon with adiabatic physics by \cite{ud-Doula2008} and \cite{ud-Doula2009} and ultimately 3D simulations of the star $\theta^{1}$ Ori C, incorporating optically thin cooling by \cite{ud-Doula2013}. More recently 2D simulations investigating the incompatibility of large scale stellar magnetic fields and the circumstellar disk found around classical Be stars have been conducted by \cite{ud-Doula2018} demonstrating that large scale fields of the order $\sim 100 \ \mathrm{G}$ will disrupt any Keplerian disk close to the star. Magnetic fields of this order in massive stars are are mostly undetected and the majority of the MIMES sample possess fields of greater strength than this \citep{Petit2013}.

The work presented here builds on the earlier work by extending the studied wind region out beyond the stellar magnetosphere to regions where the dynamic influence of the magnetic field has diminished. More crucially we also allow magnetic obliquity of the dipole, resulting in non-spherical and non-cylindrical symmetric wind evolution. We achieve this through the use of direct numerical MHD simulations. We analyse the resulting wind structure in the context of diagnosis methods developed by \cite{ud-Doula2002, ud-Doula2008, ud-Doula2009, ud-Doula2013, Petit2013}, quantify the departure from spherical symmetry over time and radial distance and finally calculate the synthetic radio lightcurves and continuum spectra; placing the results into what is observably possible with current technologies such as the JVLA and ALMA.

\subsection{O-Star thermal radio emission}

Radio emission from massive stars has historically been the subject of considerable interest \citep{Braes1972, Wright1974, Cohen1975, Panagia1975, Wright1975} as the observed emission from Plank blackbody curve calculations deviate from what is expected \citep[WB75 here after]{Wright1975}. This additional emission is due to free-free interactions in the wind. The emission is known as thermal as the wind temperature is high enough to ionise the species.

Early analytical modelling of the winds of massive stars and the calculation of radio emission from the symbiotic nova V1016 Cyg was accomplished by \cite{Seaquist1973}. The model they developed was based on the assumptions of a uniform, spherically symmetric outflow at constant temperature. The resulting spectral flux density as a function of frequency takes the form $S_{\nu} \propto \nu^{\alpha}$, where the spectral index, $\alpha$, lies between $-0.1 \leq \alpha \leq +2$.

Refinement of the model by WB75 leads to a spectral index of $\alpha = 0.6$ at radio and infrared wavelengths. The relationship between $S_{\nu}$ and $\alpha$ arises due to the optical depth of the circumstellar material possessing a different value depending on the frequency of the emission; hence for higher frequencies, emission originates from deeper within the gas and therefore a greater extent of gas contributes to the emission at that frequency, leading to this positive gradient. This concept will be covered in subsequent sections.

The precise dependence of the spectral flux on $\nu$ allows for calculating the rate at which the star is losing mass through its wind. As such, thermal radio observations provide an important window onto stellar evolution and the impact which the star has on the interstellar medium.

In \cite{Daley-Yates2016} we conducted a theoretical study of the thermal radio/sub-mm emission from a range of representative non-magnetised O-stars; with an emphasis on modulation of the continuum spectra by wind acceleration close to the stellar surface, observably accessible thanks to the sub-mm bands of ALMA. We continue this theme by applying the same analysis to the winds of magnetic massive stars.

At higher than radio/sub-mm frequencies, the observational consequences of magnetically confined winds at have been considered by \cite{Shenar2017}. They conclude that, due to the magnetic field of the star, wind material builds up leading to a density enhancement, observations of which can be exploited to determine mass-loss rates of both magnetic and non-magnetic massive stars.

\section{Modelling}

The model outlined here follows closely the methods used by \cite{ud-Doula2002} and specifically \cite{ud-Doula2013} who performed the first 3D numerical modelling of magnetised O-star winds by simulating the star $\theta^{1}$ Ori C under the adiabatic regime with optically thin radiative cooling. We deviate from this treatment by restricting our models to isothermal behaviour. The additional complication of the oblique magnetic dipole warrants this simplification as the numerical influence of the polar axis singularity becomes non-negligible (this issue will be discussed in Section \ref{sec:BCs}). The following section details the calculations and numerical schemes used to perform this simulation.

\subsection{Magnetohydrodynamics}

The winds of massive stars are accelerated to supersonic speeds within a fraction of a stellar radius, making them ideally suited to modelling via the equations of Magnetohydrodynamics (MHD):
\begin{equation}
\label{eq:mass}
\frac{\partial \rho}{\partial t} + \bm{\nabla} \cdot \left( \rho \bm{v} \right) = 0
\end{equation}
\begin{equation}
\label{eq:momentum}
\frac{\partial \bm{v}}{\partial t} + \left( \bm{v} \cdot \bm{\nabla} \right) \bm{v}
+ \frac{1}{4 \pi \rho} \bm{B} \times  \left( \nabla \times \bm{B} \right)
+ \frac{1}{\rho} \nabla p = \bm{g} + \bm{g}_{\mathrm{L}} + \bm{F}_{\mathrm{co}}
\end{equation}
\begin{equation}
\label{eq:magnetic}
\frac{\partial \bm{B}}{\partial t}
+ \nabla \times \left( \bm{B} \times \bm{v} \right) = 0.
\end{equation}
Where $\rho$, $\bm{v}$, $\bm{B}$, $p$, $\bm{g}$ and $\bm{F}_{\mathrm{co}}$ are, the density, velocity, magnetic field, pressure, acceleration due to gravity and acceleration due to the co-moving frame respectively. The additional acceleration term, $\bm{g}_{\mathrm{L}}$, describes the acceleration due to line absorption (see Section \ref{sec:radiative_driving}). $\bm{F}_{\mathrm{co}}$ is the sum of both the centrifugal and Coriolis forces: $\bm{F}_{\mathrm{co}}~=~\bm{F}_{\mathrm{centrifugal}}~+~\bm{F}_{\mathrm{coriolis}}$ which are;
\begin{equation}
\bm{F}_{\mathrm{centrifugal}} =
- \left[ \bm{\Omega}_{\mathrm{fr}} \times \left( \bm{\Omega}_{\mathrm{fr}}
\times \bm{R} \right) \right]
\end{equation}
and
\begin{equation}
\bm{F}_{\mathrm{coriolis}} =
- 2 \left( \bm{\Omega}_{\mathrm{fr}} \times \bm{v} \right)
\end{equation}
Here $\bm{\Omega}_{\mathrm{fr}}$ is the angular frequency of the rotating frame with $\bm{r}$ the radial distance vector.

As the simulation is isothermal, we close equations (\ref{eq:mass} - \ref{eq:magnetic}) using the relation:
\begin{equation}
p = \rho c_{\mathrm{iso}}^{2},
\end{equation}
in which $c_{\mathrm{iso}}$ is the isothermal sound speed, given by $c_{\mathrm{iso}}^{2}~=~2 k_{\mathrm{B}} T / m_{\mathrm{p}}$, where $k_{\mathrm{B}}$ is the Boltzmann constant, $T_{\mathrm{eff}}$ the stellar surface effective temperature and $m_{\mathrm{p}}$ the proton mass.

\subsubsection{Radiative driving}
\label{sec:radiative_driving}

The winds of massive stars are accelerated by scattering of the stellar radiation in absorption lines of elements within the wind; as such they are known as line driven winds. The theory of line driving was first established in a seminal paper by \citet[CAK here after]{Castor1975}. The principle result is the description of an expanding wind whose acceleration is governed by the local density and, under the approximation made by \cite{Sobolev1960}, the velocity gradient and is given by:
\begin{equation}
\label{eq:line_accel}
g_{L} = \frac{f_{\mathrm{D}}}{\left(1-\alpha \right)} \frac{ \kappa_{e} L_{*} \overline{Q} }{ 4 \pi r^{2} c } \left( \frac{ \mathrm{d} \bm{v} / \mathrm{d} r }{\rho c \overline{Q} \kappa_{\mathrm{e}}} \right)^{\alpha}.
\end{equation}
Where $L_{\ast}$ is the stellar luminosity, $c$ the speed of light, $\kappa_{\mathrm{e}}$ the electron scattering opacity, $\alpha$ the CAK exponent and $\bm{v}$ and  $\rho$ have the above meanings. All the parameters of equation (\ref{eq:line_accel}) are derivable from observations except for $\overline{Q}$ for which \cite{Gayley1995} computed a value of $\sim~10^{3}$ for a range of stellar parameters. The above variables are detailed in Table \ref{tab:parameters}.

Finally $f_{\mathrm{D}}$ is the finite disk correction factor, which accounts for the finite size of the stellar disk close to the star and is given by:
\begin{equation}
\label{eq:FD}
f_{\mathrm{D}} = 1 - \frac{\alpha}{2 r^{2}} \left( 1 - \frac{v_{r}}{\frac{\mathrm{d}v_{r}}{\mathrm{d}r} r} \right) \left( 1 + \frac{1 - \alpha}{2 r^{2}} \left( 1 - \frac{v_{r}}{\frac{\mathrm{d}v_{r}}{\mathrm{d}r} r} \right) \right)
\end{equation}
(private communication Stan Owocki). Where $\mathrm{d} v_{r} / \mathrm{d} r$ is the gradient of the radial component of velocity in the radial direction, non-radial acceleration is neglected here.

Non-radial contributions to the radiative acceleration become important when considering fast rotating stars \citep{Gayley2000}. As the rotational confinement parameter $W$ is relatively modest as $~11\%$ of critical, we consider non-radial acceleration as negligible. For more complete considerations of non-radial radiative driving see the recent works by \cite{Pittard2009}, \cite{Kee2016}, \cite{Sundqvist2018} and \cite{Owocki2018}.



Equation (\ref{eq:line_accel}) is applied on the right hand side of equation (\ref{eq:momentum}) as a source term alongside gravitational and rotational source terms.

\subsection{Stellar parameters}

For the simulated star, we take parameters from \cite{Daley-Yates2016}, whose stellar models are derived from the data of \cite{Krticka2014}. We use model S3 from the former work and summarise the parameters in Table \ref{tab:parameters}.

The mass-loss rate, $\dot{M}_{B=0}$ refers to the mass-loss rate of a star with the same parameters but with no magnetic field and is calculated, according to \cite{Owocki2004}, in the following manner:
\begin{equation}
\dot{M}_{B=0} = \frac{L_{\ast}}{c^{2}} \left( \frac{\alpha}{1 - \alpha} \right) \left( \frac{\overline{Q} \Gamma_{\mathrm{e}}}{1 - \Gamma_{\mathrm{e}}} \right)^{\frac{1 - \alpha}{\alpha}} \left( 1 + \alpha \right)^{-1/\alpha}.
\end{equation}
Where $L_{\ast}$, $c$, $\alpha$ and $\overline{Q}$ have their previous meanings and $\Gamma_{\mathrm{e}}$ is the Eddington parameter.

We use $\dot{M}_{B=0}$ to specify the initial conditions of the density profile via the expression
\begin{equation}
\label{eq:density_profile}
\rho = \frac{\dot{M}_{B=0}}{4 \pi r^{2} v(r)}
\end{equation}
where the velocity profile is
\begin{equation}
\label{eq:velocity_profile}
v(r) = v_{\infty} (1 - R_{\ast}/r)^{\beta},
\end{equation}
with $\beta$ determining the steepness of the velocity profile and $v_{\infty}$ is the wind terminal velocity. The mass-loss from the star in the simulation deviates from this idealised value due to confinement by the magnetic field, the actual mass-loss rate is measured from the simulation results.

Instead of specifying the equatorial field strength directly, we specify the dimensionless magnetic field confinement parameter to have a value $\eta_{\ast}~=~20$, resulting in a equatorial magnetic field, $B_{\mathrm{eq}}~=~324 \ \mathrm{G}$ (see equation (\ref{eq:eta}), of the following section). This value was chosen as a balance between what is numerically feasible and physically representative, based on data from the MIMES project \citep{Petit2013, Wade2016}. The value is similar in magnitude to the O-type stars HD 108 and NU Ori and the B-type stars HD~66665 and $\sigma$ Lup.

Larger equatorial magnetic field strengths lead to more restrictive numerical time steps. As such, $\eta_{\ast}~=~20$, produces the desired behaviour, in perturbing the stellar wind to form an excretion disk but allows for the simulation to be run in a feasible time span.

The angle of magnetic field obliquity, $\zeta$, is constrained by the presence of the polar boundaries at $\theta~=~0$ and $\theta~=~\pi$. This issue will be discussed in more depth in Section \ref{sec:BCs}, however it is necessary to state here that the boundary restricts the obliquity of the dipole and that greater obliquity leads to enhanced numerical effects at the aforementioned boundaries. The chosen value of $\zeta~=~30^{\circ}$ reflects this issue and was deemed a significant enough obliquity to promote the desired perturbation to the stellar wind, yet small such that numerical effects are kept to a negligible level.

The remaining parameters in Table \ref{tab:parameters} are used to calculate equation (\ref{eq:line_accel}) and to parametrise the simulation code units. For example, in the results in Section \ref{sec:results}, distances are given in stellar radii.

\begin{table}
	\caption[Stellar parameters for the simulated O-star.]{Stellar parameters for the study. \label{tab:parameters}}
	\centering
	\begin{tabular}{ccc}
		\hline
		Name & Parameter & Value \\
		\hline
		Initial mass-loss & $\dot{M}_{B=0}$ & $10^{-7}$ M$_{\odot}$ yr$^{-1}$   \\
		Stellar radius & $R_{\ast}$ & 9 $R_{\odot}$ \\
		Stellar mass & $M_{\ast}$ & 26 $M_{\odot}$  \\
		Distance to observer & $D$ & 0.5 kpc \\
		Effective temperature & $T_{\mathrm{eff}}$ & 36300 K \\
		Luminosity & $\log_{10} (L_{\ast}/L_{\odot})$ & 5.06 \\
		Eddington factor & $\Gamma_{Edd}$ & 0.11 \\
		Q-factor & $\overline{Q}$ & 700 \\
		Escape velocity & $v_{\mathrm{esc}}$ & 1000 km s$^{-1}$ \\
		Terminal velocity & $v_{\infty}$ & 1228 km s$^{-1}$ \\
		Rotational velocity & $v_{\mathrm{rot}}$ & 82 km s$^{-1}$ \\
		Keplerian orbital speed & $v_{\mathrm{orb}}$ & 751 km s$^{-1}$ \\
		Rotational rate & $\omega$ & 0.2 $\omega_{\mathrm{crit}}$ \\
		CAK exponent & $\alpha$ & 0.6 \\
		Velocity law & $\beta$ & 0.8 \\
		Magnetic field inclination & $\zeta$ & $30^{\circ}$ \\
		Rotation parameter & $W$ & 0.11 \\
		Confinement parameter & $\eta_{\ast}$ & 20.0 \\
		Kepler radius & $R_{\mathrm{K}}$ & 4.56 $R_{\ast}$ \\
		Alfv\'{e}n radius & $R_{\mathrm{A}}$ & 3.98 $R_{\ast}$ \\
		\hline
	\end{tabular}
\end{table}

\subsection{Magnetosphere characterisation}
\label{sec:magnetosphere}

\cite{Petit2013} presents a scheme which characterises the global behaviour of a massive star magnetosphere as a function of several dimensional quantities developed by \cite{ud-Doula2002} and \cite{ud-Doula2008}. These quantities are the wind magnetic confinement parameter,
\begin{equation}
\label{eq:eta}
\eta_{*} \equiv \frac{B_{\mathrm{eq}}^{2} R_{*}^{2}}
{\dot{M}_{B=0} v_{\infty}}
\end{equation}
and the ratio of the rotational speed $v_{\mathrm{rot}}$
to the Keplerian orbital speed $v_{\mathrm{orb}}$,
\begin{equation}
W \equiv \frac{v_{\mathrm{rot}}}{v_{\mathrm{orb}}} =
\frac{\omega R_{*}}{\sqrt{G M_{*} / R_{*}}}.
\end{equation}
In the above expressions, $B_{\mathrm{eq}}$ is the equatorial magnetic field strength, $R_{\ast}$ is the stellar radius, $\dot{M}_{B=0}$ is the stellar mass-loss rate in the presence of no magnetic field and $v_{\infty}$ is the wind terminal velocity. $\omega$, $G$ and $M_{\ast}$ are the angular rotational frequency, gravitational constant and stellar mass respectively.
The Kepler and Alfv\'{e}n radii are then calculated from
\begin{equation}
R_{\mathrm{K}} = W^{-2/3} \left[R_{\ast}\right]
\label{eq:radius_kepler}
\end{equation}
and
\begin{equation}
R_{\mathrm{A}} = 0.3 + (\eta_{\ast} + 0.25)^{1/4} \left[R_{\ast}\right].
\label{eq:radius_alfven}
\end{equation}
See \cite{Petit2013} for a detailed discussion of these expressions.

When calculating $R_{\mathrm{A}}$ using equation (\ref{eq:radius_alfven}), a value of 2.43 $R_{\ast}$ was found, this however is an underestimate when compared to that found from the simulation results, 3.98 $R_{\ast}$. To be consistent with our analysis, when using $R_{\mathrm{A}}$, we will use the simulated value rather than the prediction. The simulated value is quoted in Table \ref{tab:parameters}.

The case when $\eta_{\ast} >> 1$ represents a strongly confined wind where the magnetic pressure dominates and conversely when $\eta_{\ast} << 1$ the field is weak and the wind ram pressure dominates. For the rotational parameter, $W = 1$ represents the critical stellar breakup rotational speed, where the gravitational acceleration equals the rotational acceleration at the stellar equator. Together the above two parameters characterise the dynamics of material suspended in the stellar magnetosphere.

\cite{Petit2013} divided massive star magnetospheres into two distinct categories; dynamical and centrifugal. The division is determined by the relative values of the Kepler and Alfv\'{e}n radii. For a star with $R_{\mathrm{K}} > R_{\mathrm{A}}$, it's magnetosphere is defined as dynamical and wind material confined on closed magnetic field lines experiences an unstable equilibrium and consequently there is continuous motion of material as field lines are loaded and emptied. However, for a star with $R_{\mathrm{K}} < R_{\mathrm{A}}$, there exists a region between the two radii in which material experiences a stable equilibrium between gravity, magnetic tension and centrifugal acceleration, below $R_{\mathrm{K}}$ behaviour is still dynamical, as $R_{\mathrm{K}} > R_{\mathrm{A}}$. Under this framework, the star in our simulation has a dynamical magnetosphere. See Table \ref{tab:parameters} for the values of $\eta_{\ast}$, $W$, $R_{\mathrm{K}}$ and $R_{\mathrm{A}}$.

\begin{figure}
	\centering
	\includegraphics[width=0.49\textwidth,trim={0cm, 3cm, 0cm, 0cm},clip]{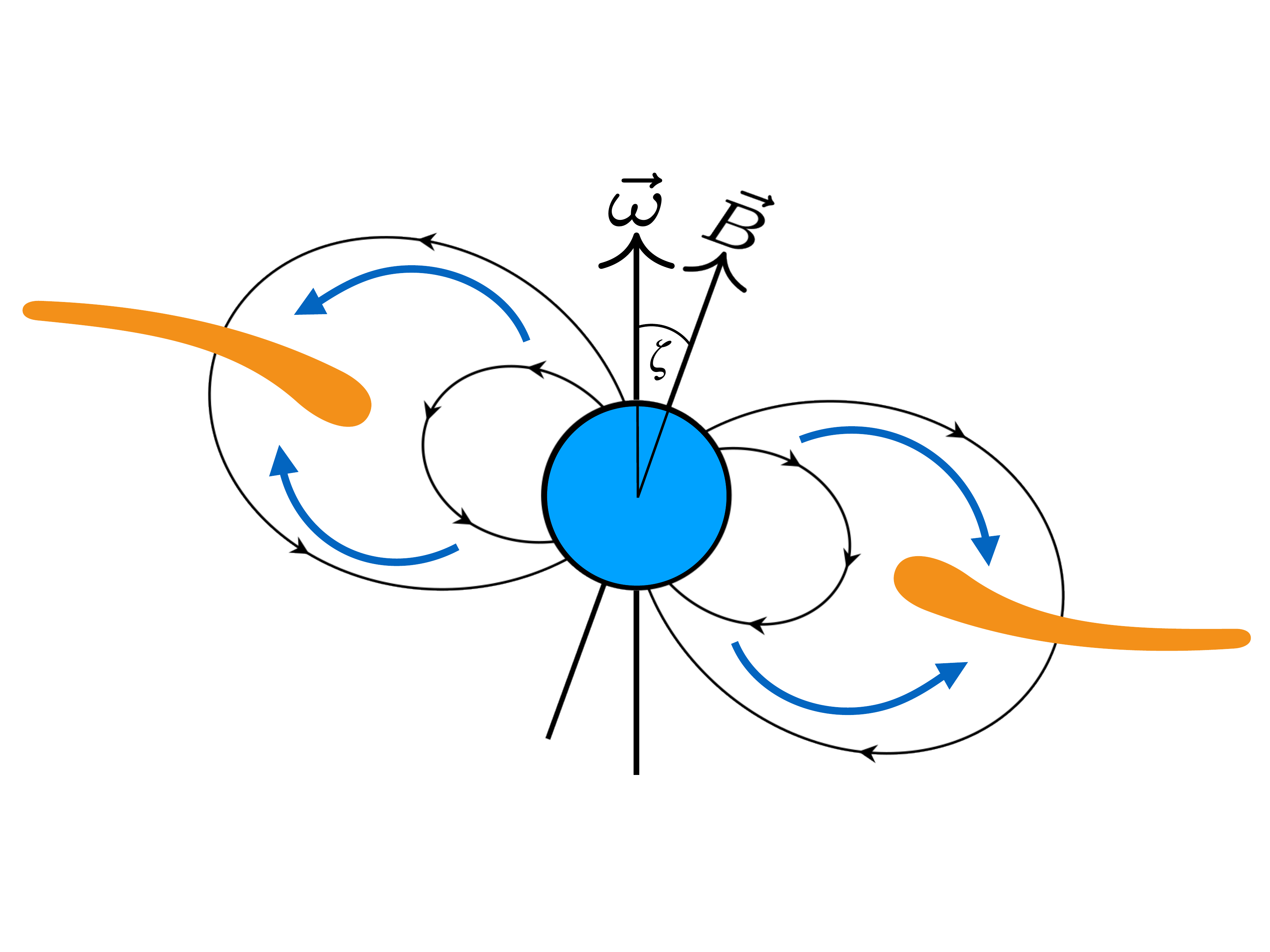}
	\caption[Cartoon diagram illustrating the magnetic field topology and confinement of the O-star wind.]{Cartoon diagram illustrating the magnetic field topology and confinement of the stellar wind. The blue circle at the centre is the star, annotated are the magnetic field lines with their vector direction and the dipole obliquity, $\zeta = 30^{\circ}$. The blue arrows indicate the path material travels along as the wind interacts with the magnetic field. The orange tear-drop like shapes represent the shocked wind material. \label{fig:diagram}}
\end{figure}

Out beyond both $R_{\mathrm{K}}$ and $R_{\mathrm{A}}$ material undergoes a net outward acceleration as $g_{\mathrm{L}}$ exceeds all other inward acceleration. As this material has already undergone confinement and perturbation away from its initial spherical surface velocity, an excretion disk-like structure develops whose outward path intersects with the apex of the closed magnetic field lines. The net effect is to produce a standing excretion disk-like shaped shock structure with a topology intimately linked to the magnetic field topology. This excretion disk-like structure appears as a spiral when viewed in 2D. Fig. \ref{fig:diagram} illustrates this interplay between outward flow of the wind and its confinement by the dipole magnetosphere. As can be seen in profile in Fig. \ref{fig:diagram}, for an oblique dipole, the shock structure forms a contorted disk.

The following section explains the procedure for calculating the synthetic thermal radio emission from the structure described above.

\subsection{Synthetic radio/sub-mm emission}
\label{sec:theory_radio}

The numerical procedure for calculating synthetic radio emission follows the one we developed in \cite{Daley-Yates2016}. For the present study it will suffice to cover the equations directly used in the calculation and we direct the interested reader to the aforementioned paper for a full description of the theory.

Recently, \cite{Fionnagain2018} and \cite{Kavanagh2019} have employed a very similar method for calculating radio emission but for the case of solar mass stars.

The specific intensity of radio emission for each column along the line of sight from the observer through the simulation domain is given by:
\begin{equation}
\label{eq:I_iso}
I_{\nu}(y,z) = B_{\nu} \left( T \right) \int_{0}^{\tau_{\mathrm{max}}(y,z)} \exp(-\tau (x,y,z)) \mathrm{d} \tau,
\end{equation}
where $B_{\nu}(T)$ is the Planck function at frequency $\nu$, $T$ the temperature, $\tau$ the optical depth and $\tau_{\mathrm{max}}$ is the maximum optical depth along the observer's line of sight. Equation (\ref{eq:I_iso}) is integrated to give
\begin{equation}
\label{eq:I_tau_max}
I_{\nu}(y,z) = B_{\nu} \left( T \right) \left[1-\exp(-\tau_{\mathrm{max}}(y,z)) \right].
\end{equation}
Where
\begin{equation}
\label{eq:tau_continus}
\tau_{\mathrm{max}}(y,z) = \int_{-\infty}^{+\infty} \gamma K_{\nu}(T_{\mathrm{eff}}) n_{\mathrm{i}}^{2}(x,y,z) \mathrm{d} x.
\end{equation}
$n_{\mathrm{i}}$ is the ion number density and $\gamma$ is the ratio of the electron and ion number densities, $\gamma~=~1.01$, under the assumption of solar metallicity. The final variable in equation (\ref{eq:tau_continus}) is
\begin{equation}
\label{eq:kappa_nu_t}
K_{\nu}(T_{\mathrm{eff}}) = 0.0178 \frac{Z^{2} \textit{\textg}_{\mathrm{ff}} }{T_{\mathrm{eff}}^{3/2} \nu^{2}},
\end{equation}
which relates the temperature, $T_{\mathrm{eff}}$, metallicity, $Z$, observing frequency, $\nu$ and \textit{\textg}$_{\mathrm{ff}}$, which is the free-free Gaunt factor, given by:
\begin{equation}
\mathrm{\textit{\textg}}_{\mathrm{ff}} = 9.77 + 1.27 \log_{10} \left( \frac{T_{\mathrm{eff}}^{3/2}}{\nu Z} \right),
\label{eq:gff}
\end{equation}
in which the symbols have their above meaning \citep{stevens1995, Hoof2014, Daley-Yates2016}. Finally, the total spectral flux emitted at a frequency, $\nu$, is then the integral of the specific intensity $I_{\nu}$ over the $yz$-plane:
\begin{equation}
\label{eq:flux_contiuns}
S_{\nu} = \frac{1}{D^{2}} \int_{0}^{\infty} \int_{0}^{\infty} I_{\nu}(y,z) \mathrm{d}y \mathrm{d}z,
\end{equation}
with $D$ the distance between the observer and the star, kept at a constant value of $0.5 \ \mathrm{kpc}$ throughout the calculations, and $I_{\nu}$ given by equation (\ref{eq:I_tau_max}).

Equations (\ref{eq:I_tau_max}, \ref{eq:tau_continus} and \ref{eq:flux_contiuns}) are the primary expressions used for calculating the synthetic radio emission, the results of which are presented in Section \ref{sec:emission_res}.

It is possible to determine the mass-loss rate, $\dot{M}_{\ast}$, from the free-free radio emission of massive stars. In a seminal paper, WB75 construct a framework for predicting the spectral flux, $S_{\nu}$ as a function of $\dot{M}_{\ast}$. Their equation (equation (8) of WB75) can therefore be algebraically manipulated to give $\dot{M}_{\ast}$ in terms of the stellar properties, listed in Table \ref{tab:parameters}, and $S_{\nu}$. The exact expression is given by \cite{Bieging1989} as:
\begin{equation}
\dot{M}_{\mathrm{obs}} = \frac{3.01 \times 10^{-6} \mu}{Z(\gamma \mathrm{\textit{\textg}}_{\mathrm{ff}} \nu)^{1/2}} v_{\infty} S_{\nu}^{3/4} D^{3/2	} M_{\odot} yr^{-1}.
\label{eq:WB_M_dot}
\end{equation}
Where $\mu$ is the mean atomic weight, $v_{\infty}$ is in km/s and $D$ is in kpc. The remaining variables are in cgs units.


\subsection{Simulation}

The MHD equations (\ref{eq:mass} - \ref{eq:magnetic}) were solved using the publicly available code PLUTO (version 4.2) \citep{Mignone2007}.

The chosen algorithm was fully unsplit and 2nd order accurate in space and time, using linear reconstruction, Runge-Kutta time stepping and employed the HLL Riemann solver. The extended GLM divergence cleaning algorithm was used to ensure the $\nabla \cdot \bm{B} = 0$ condition.

\subsubsection{Numerical grid}

The numerical grid in our simulation covered a physical extent of $r \in \{1, 40\} \ R_{\ast}$, $\theta \in \{0, \pi\} \ \mathrm{radians}$ and $\phi \in \{0, 2 \pi\} \ \mathrm{radians}$. This provided a computational region extending from the stellar surface to the outer wind, far beyond the magnetospheric radius, thus facilitating the capture of low frequency radio emission generated by the  extended wind.

The simulations were performed using a stretched rectilinear spherical polar grid in which the physical volume was discretised with 300 cells in $r$, 120 cells in $\theta$ and 240 cells in $\phi$. This leads to a cell size in the $r$ direction which stretched from $\Delta r_{1}~\approx~0.0007 \ R_{\ast}$ to $\Delta r_{300}~\approx~0.93 \ R_{\ast}$ with a constant stretching factor of $1.0243$. Both the $\theta$ and $\phi$ directions have equal spacing of $\Delta \theta_{j}~=~\Delta \phi_{k}~\approx~0.026 \ \mathrm{radians}$. The stretching regime in the radial direction is required to resolve the sonic point of the wind, which for our simulated star lies at $1.0018 \ R_{\ast}$.

\subsubsection{Initial conditions}
\label{sec:IC}

The initial conditions of the simulations were specified using the density and velocity profile equations (\ref{eq:density_profile}) and (\ref{eq:velocity_profile}).

The magnetic field was initialised as a perfect dipole, centred at the origin and rotated about the $y$-axis, in the $xz$-plane. This configuration then relaxes to quasi-steady state as the simulation evolves.

\subsubsection{Boundary conditions}
\label{sec:BCs}

The outer radial boundary of the simulation is set to outflow. The inner radial boundary is set such that the the star is continually feeding material to the wind and therefore replenishing material in the simulation. As the wind is accelerated to supersonic radial speeds within a fraction of a stellar radii, and as the line driving is dependent on the velocity gradient (equation (\ref{eq:line_accel})), the evolution of the entire simulation is a boundary value problem which is sensitively dependent on the lower radial boundary condition. To account for this sensitivity we used the boundary conditions of \cite{ud-Doula2002} and \cite{ud-Doula2013}. The density is specified via equation (\ref{eq:velocity_profile}), replacing the velocity profile with a ratio linked to the sound speed; $\rho~=~\dot{M} / 4 \pi R^{2}_{\ast} (c_{\mathrm{iso}}/\xi) $ where $\xi$ is a factor parametrised to give a stable material inflow at the boundary and is typically $5~<~\xi~<~30$. Values of $\xi$ outside this range can result in oscillations of the solution at the boundary.

The velocity in the lower radial boundary is specified by linearly extrapolating back from the first 2 computational cells above the boundary, allowing the flow into the computational active zone to adjust to the conditions of the wind and permitting material to also re-enter the stellar surface as magnetically confined material follows field lines back to the stellar surface. Specifying the boundary in this manner also allows the mass loading of the wind to self consistently adapt to the rotation of the star. Large rotational velocities can impact the mass-loss of a star. This is due to the effective gravity at the rotational equator being reduced relative to the poles, leading to material being lifted from the surface more easily.

The boundary of the lower and upper azimuthal direction is made reciprocal such that material can move freely around the star. The upper and lower boundary of the polar direction was set to reflective so as not to act as a sink for material. This final boundary condition is non-physical and a reflective polar boundary can lead to spurious heating along the polar axis. There are several methods designed to overcome this numerical difficulty. One such method is known as $\pi$-boundary conditions in which the fluid quantities are translated $\pi$ around the axis and vector values transformed such that material effectively passes over the pole. This method is implemented in the public codes Athena++ \citep{White2016} and MPI-AMRVAC \citep{Xia2018}. PLUTO does not provide this functionality however. Another means of avoiding the spurious heating is to average fluid quantities over the poles and effectively smooth over the anomalies.

As the current study aim is to quantify the thermal radio emission and as this emission is weakly dependent on the gas temperature ($\propto T^{0.11}$), we have chosen to run the simulation under an isothermal equation of state. This effectively side steps the issue of polar boundary heating, as the temperature is constant at $T_{\mathrm{eff}}$ and not evolved with time. The extent to which the boundary impacts the density, velocity and magnetic profiles is discussed in Section \ref{sec:structure}.

The isothermal assumption forces us to neglect behaviour due to both shock heating and radiative cooling. Both of which have been shown to play a role in the wind dynamics \citep{ud-Doula2008, ud-Doula2013}. As such, this is a limitation of the present study and fully adiabatic simulations with cooling physics are the aim of future studies.

\subsubsection{Steady state criteria}

The simulation were deemed to have reached steady state after a simulation time of 1~Ms or approximately 20 stellar rotations. This time frames allows material to relax from the initial spherical symmetry to the confined configuration. Any excess material is also blown off to the outer boundary within this time frame.

Once the 1 Ms had been reached, we ran the simulation for a further 2 rotations with fine time spaced sampling. This allowed for high temporal resolution for the synthetic radio light curve calculations in Section \ref{sec:rotation}.

\section{Results and discussion}
\label{sec:results}

The following subsections layout the simulation results; starting with the global simulation properties and profiles for the three primary fluid quantities: $\rho$, $|\bm{v}|$ and $|\bm{B}|$. We will then quantify the extent to which the wind has deviated from spherical symmetry and finally examine both the synthetic radio lightcurves and the radio/sub-mm spectrum.

\subsection{Global properties}

Next we will cover the issue of convergence and ascertain whether quasi-steady state has been reached.

\subsubsection{Mass and angular momentum flux}
\label{sec:comparison}

\begin{figure}
	\centering
	\includegraphics[width=0.49\textwidth]{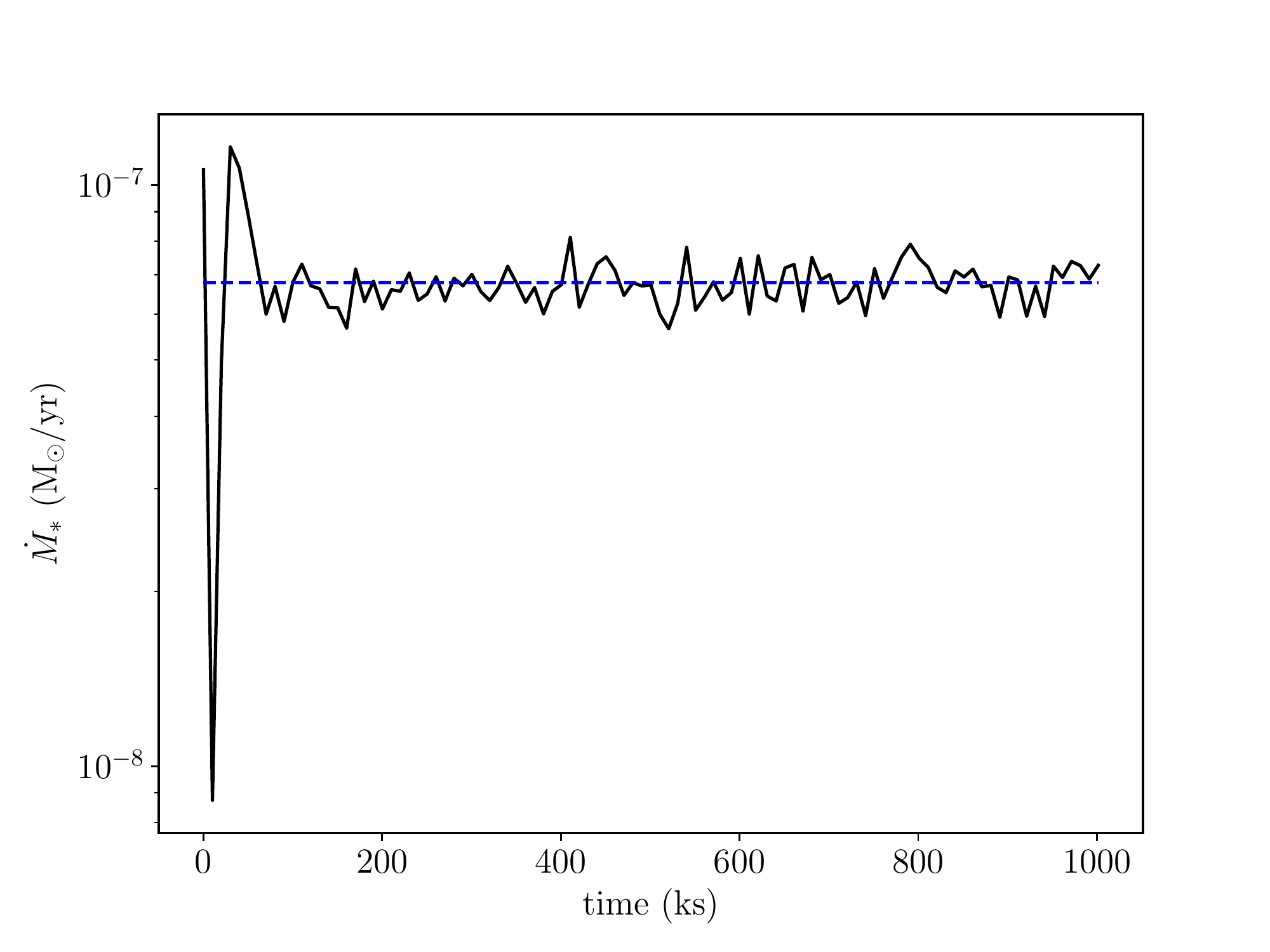}
	\caption[Evolution of the mass-loss rate of the simulated O-star.]{Evolution of the mass-loss over the course of the simulation. The mass flux from the star is initially $\sim~10^{-7} \ \mathrm{M_{\odot}/yr}$ and relaxes to an average value of $6.8~\times~10^{-8} \ \mathrm{M_{\odot}/yr}$ (indicated by the blue dashed line) after $\sim~100 \ \mathrm{ks}$ where it remains for the rest of the simulation.
		\label{fig:mass_loss}}
\end{figure}

Fig. \ref{fig:mass_loss} shows the evolution of the mass-loss, $\dot{M}_{\ast}$. The initial mass-loss, $\sim~10^{-7} \ \mathrm{M_{\sun}/yr}$ undergoes a drastic reduction and oscillation as the spherical wind reacts to the presence of the magnetic field. This initial phase then stabilises to an average mass-loss of $6.8~\times~10^{-8} \ \mathrm{M_{\sun}/yr}$ after $\sim~100 \ \mathrm{ks}$.

There is still an oscillation amplitude of $\sim~1.5\%$ about this average value due to motion of material suspended on closed field lines within the inner magnetosphere. As magnetic tension, gravity, centrifugal and radiative acceleration balance in unstable equilibrium, material can either leave or re-enter the stellar surface, resulting in the $\dot{M}_{\ast}$ oscillations.

\begin{figure*}
	\centering
	\includegraphics[width=1.0\textwidth,trim={1.5cm 3cm 1.5cm 3cm},clip]{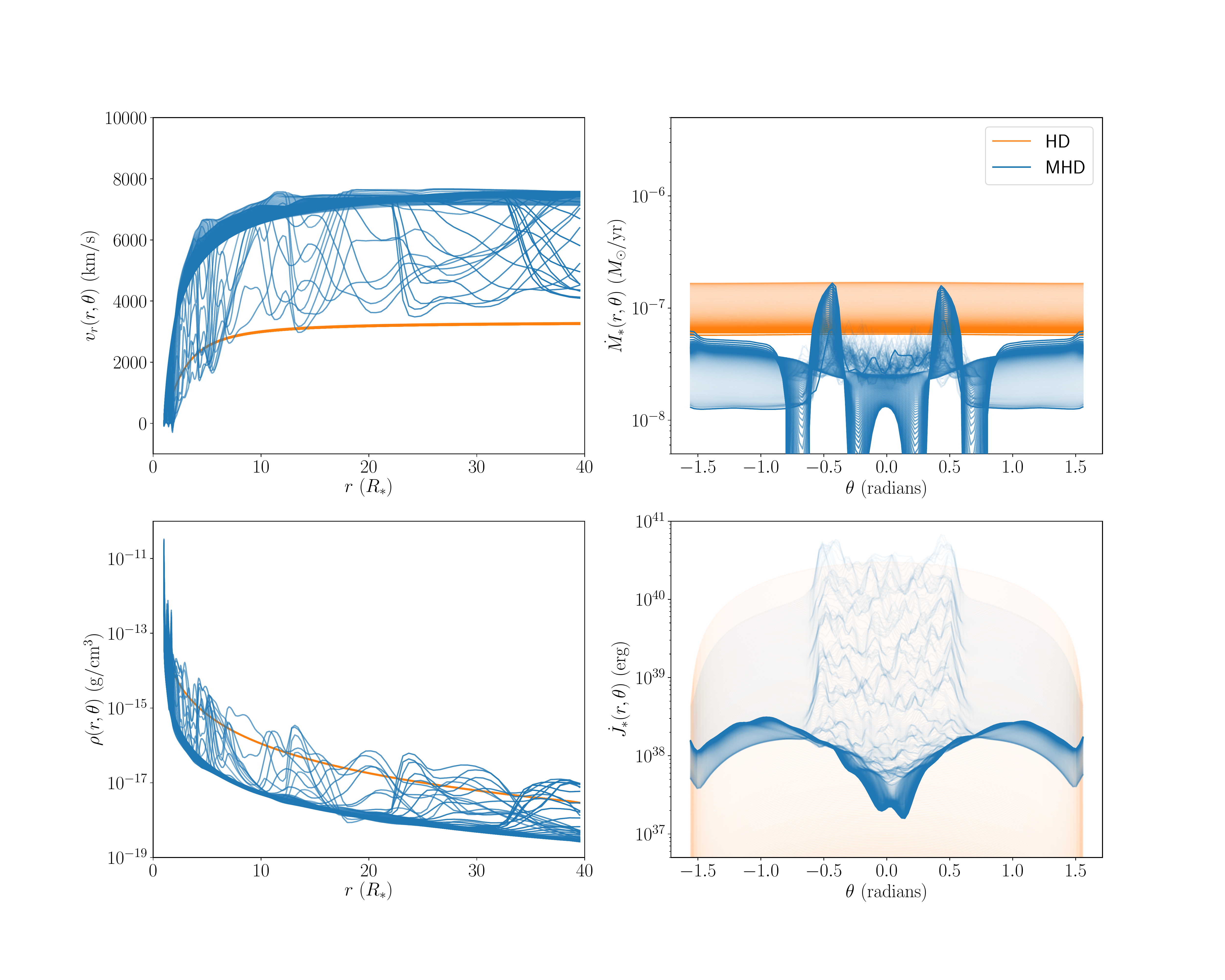}
	\caption{Comparison between MHD and HD models. Left column, radial velocity and density as a function of radial distance from the stellar surface, each curve represents a different value of $\theta$ with the opacity of each curve denoting its distance from the rotational equator $\theta~=~0$, fainter curves being further from the equator, both plots are at $\phi~=~0$. Right column, mass-loss rate and angular momentum flux as a function of $\theta$, each line represents a different value of $r$ with the opacity increasing with distance from the stellar surface, the quantities are averaged in the azimuthal direction. The broad region of enhanced mass-loss and angular momentum flux in the rotational equator is indicative of the excretion disk, highlighting the channeling of material by the magnetic field. \label{fig:comparison}}
\end{figure*}

To gain a clearer picture of the steady state wind and to benchmark the magnetised wind against a simpler, purely hydrodynamic wind, we calculate the mass-loss and angular momentum flux as a function of $\theta$, for successive radial distances. For both quantities, This is done by calculating the point values at every cell at a given radius and then multiply by the area of the sphere at that radius, effectively making each value correspond to an isolated, independent measure of the global mass-loss rate, or angular momentum flux. As we shall see below, both rotation and magnetic confinement break the symmetry of the wind and show two distinct modes; the free streaming wind and a magnetically confined disk region. The two expressions are, for the mass-loss rate,
\begin{equation}
\label{eq:mass_loss}
\dot{M}_{\ast} (r, \theta) = 2 \pi r^{2} v_{r}(r, \theta) \rho(r, \theta),
\end{equation}
and for the angular momentum,
\begin{equation}
\label{eq:am_loss}
\begin{aligned}
\dot{J}_{\ast} (r, \theta) = 2 \pi r^{2} r_{\mathrm{cy}}(r, \theta) \Bigg( v_{\phi}(r, \theta) v_{r}(r, \theta) \rho(r, \theta) \\
- \frac{B_{\phi}(r, \theta) B_{r}(r, \theta)}{4 \pi} \Bigg).
\end{aligned}
\end{equation}
Where $r_{\mathrm{cy}}(r, \theta)$ is the cylindrical radial distance from the $z$-axis \citep{Vidotto2014a, Usmanov2018}. These quantities, together with the radial velocity and density profiles are plotted in Fig. \ref{fig:comparison} at 1000 ks. The MHD results (blue curves) are plotted alongside the results for a non-magnetised HD version of the simulation (orange curves) for comparison and to highlight the departure from spherical symmetry. The mass-loss rate displays an overall reduction, with the majority of material confined to flow in the region $-0.6 \ \mathrm{radians}~<~\theta~<~0.6 \ \mathrm{radians}$ either side of the rotational equator. This is consistent with what is expected due to magnetic confinement of the wind material and the moderate extent of the dipole obliquity. Broadening of this region would occur if the obliquity angle were to be increased, and vice versa if reduced. The angular momentum loss rates follow a similar confinement pattern, however $\dot{J}$ is reduced to a less degree and in the equatorial region undergoes an increase with respect to the non-magnetised case. Both the mass and angular momentum loss in the right hand column of Fig. \ref{fig:comparison} shows that our model is capable of obtaining a smooth solution when the magnetic field does not perturb the wind and that there is significant departure from this smooth solution when wind is magnetised.

The radial profiles in the left column show the largest departure from the non-magnetised solution. While the density (lower left panel) displays a profile that is consistent with the mass-loss rate result, reduced for all distances and latitudes, the radial velocity is remarkable as $v_{\infty}$ is more than twice the value of the non-magnetised case at all latitudes. When compared to $v_{\mathrm{esc}}$ it is expected that $v_{\infty}~\sim~3 v_{\mathrm{esc}}$, however we find that when the wind is magnetised, $v_{\infty}~=~7 v_{\mathrm{esc}}$, a significant increase. A $v_{\infty}$ of this order was communicated by \cite{Friend1984}, who attributed an increased $v_{\infty}$ to the deposition of momentum into the wind through magnetic and centrifugal forces. However, their analysis neglected the finite disk correction factor (equation (\ref{eq:FD})), which leads to more modest increases in $v_{\infty}$ when accounted for. The increase in $v_{\infty}$ we observe in our magnetised simulation is larger than the generally excepted value of $\sim~3v_{\mathrm{esc}}$ \citep{ud-Doula2002, Owocki2004} however, as our non-magnetised simulation reproduces this value and that the only difference is the presence of the magnetic field, we can conclude that the radiative driving is behaving as intended and make the assumption that a combination of both magnetic and centrifugal action are responsible for the increased $v_{\infty}$.

\subsubsection{Magnetic field wind modulation}

\begin{figure}
	\centering
	\includegraphics[width=0.49\textwidth]{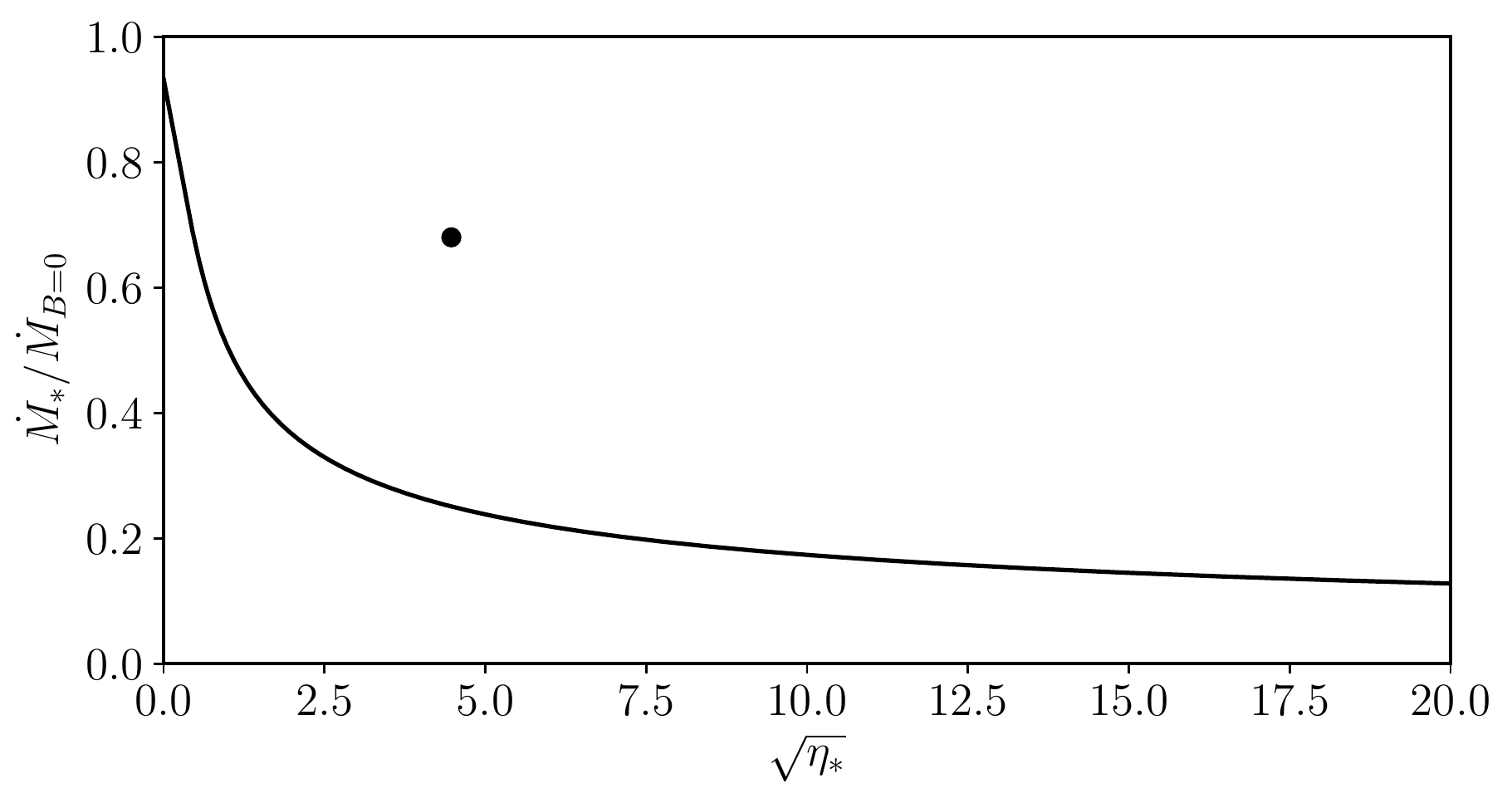}
	\caption[Mass-loss dependence on dimensionless magnetic and rotation confinement parameters.]{Dependence of the mass-loss $\dot{M}_{\ast}$ on the confinement parameter $\eta_{\ast}$ for $W = 0.11$. The black dot indicates the measured mass-loss reduction from the simulations as the ratio of the initial and averaged quasi-steady state mass-losses and has a value $\dot{M}_{\ast}/\dot{M}_{\ast, B=0}~=~0.68$.
		\label{fig:mass_loss_reduction}}
\end{figure}

\cite{ud-Doula2008} derived an expression describing the manner in which $\dot{M}_{B=0}$ is modulated by $\eta_{\ast}$ and $W$ (see equation (24) and Fig. 8 of the aforementioned paper) for a 2D axisymmteric wind. We plot this function for the stellar parameters in Table \ref{tab:parameters} and a range of $\eta_{\ast}$ values in Fig. \ref{fig:mass_loss_reduction}. The curve in this figure represents the prediction of \cite{ud-Doula2008} and the black dot, the measurment directly from our simulation and has a value of $\dot{M}_{\ast}/\dot{M}_{\ast, B=0}~=~0.68$. This result is in contrast to the predicted value of $\dot{M}_{\ast}/\dot{M}_{\ast, B=0}~=~0.25$ being approximately 2.7 times larger. This means that the mass-loss of our simulated star retains much of the equivalent non-magnetic value and is larger than predicted by \cite{ud-Doula2008}.

Our simulated star has non-aligned magnetic and rotational equators, this could potentially lead to a reversal of the effect of $\dot{M}_{\ast}$ reduction induced by the magnetic field. As this simulation is restricted to a single set of stellar parameters, we leave the effect of dipole obliquity on stellar mass-loss to a future parameter study.

As the mass-loss rate evolution in Fig. \ref{fig:mass_loss} attests, the simulation has reached quasi-steady state by $t~=~1 \ \mathrm{Ms}$, the profiles in the following section are therefore taken at this time point.

\subsection{Wind structure}
\label{sec:structure}

\begin{figure*}
	\centering
	\includegraphics[width=1.0\textwidth,trim={0cm 3cm 0cm 3cm},clip]{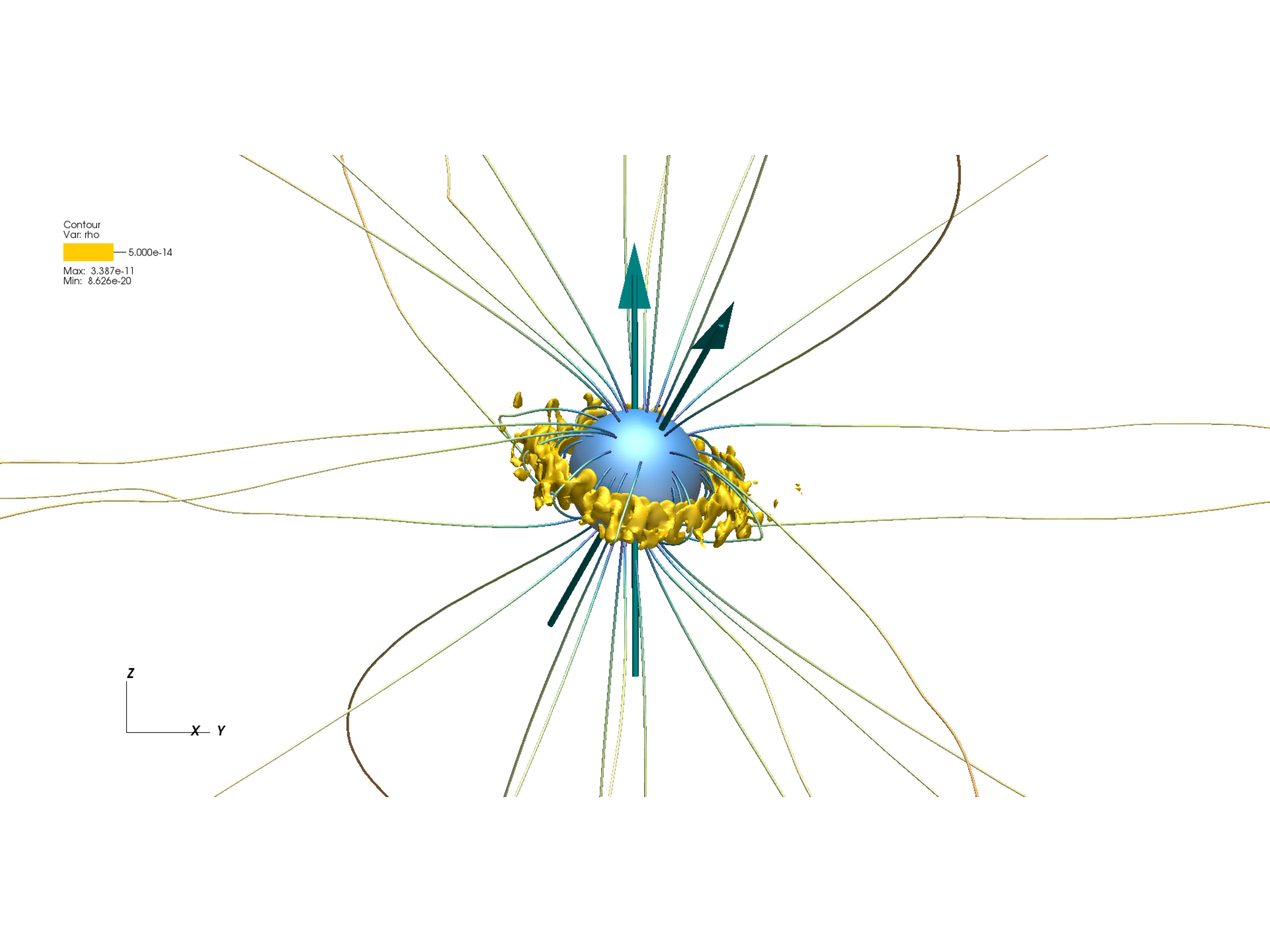}
	\caption[3D representation of the simulation domain for the immediate surroundings of the O-star.]{3D representation of the simulation domain for the immediate surroundings of the star. The blue sphere at the centre indicates the stellar surface, the streamlines are the magnetic field, coloured by the log of the magnetic field magnitude and the yellow contour is an isodensity surface of $5~\times~10^{-14} \ \mathrm{g/cm}^{3}$. While the value of this contour is some what arbitrary, it aids in illustrating the confinement of the stellar wind and the departure from spherical symmetry due to the magnetic field. The two arrows indicate the rotational axis (vertical arrow) and the magnetic dipole vector (oblique arrow). \label{fig:volume}}
\end{figure*}

\begin{figure*}
	\centering
	\includegraphics[width=1.0\textwidth]{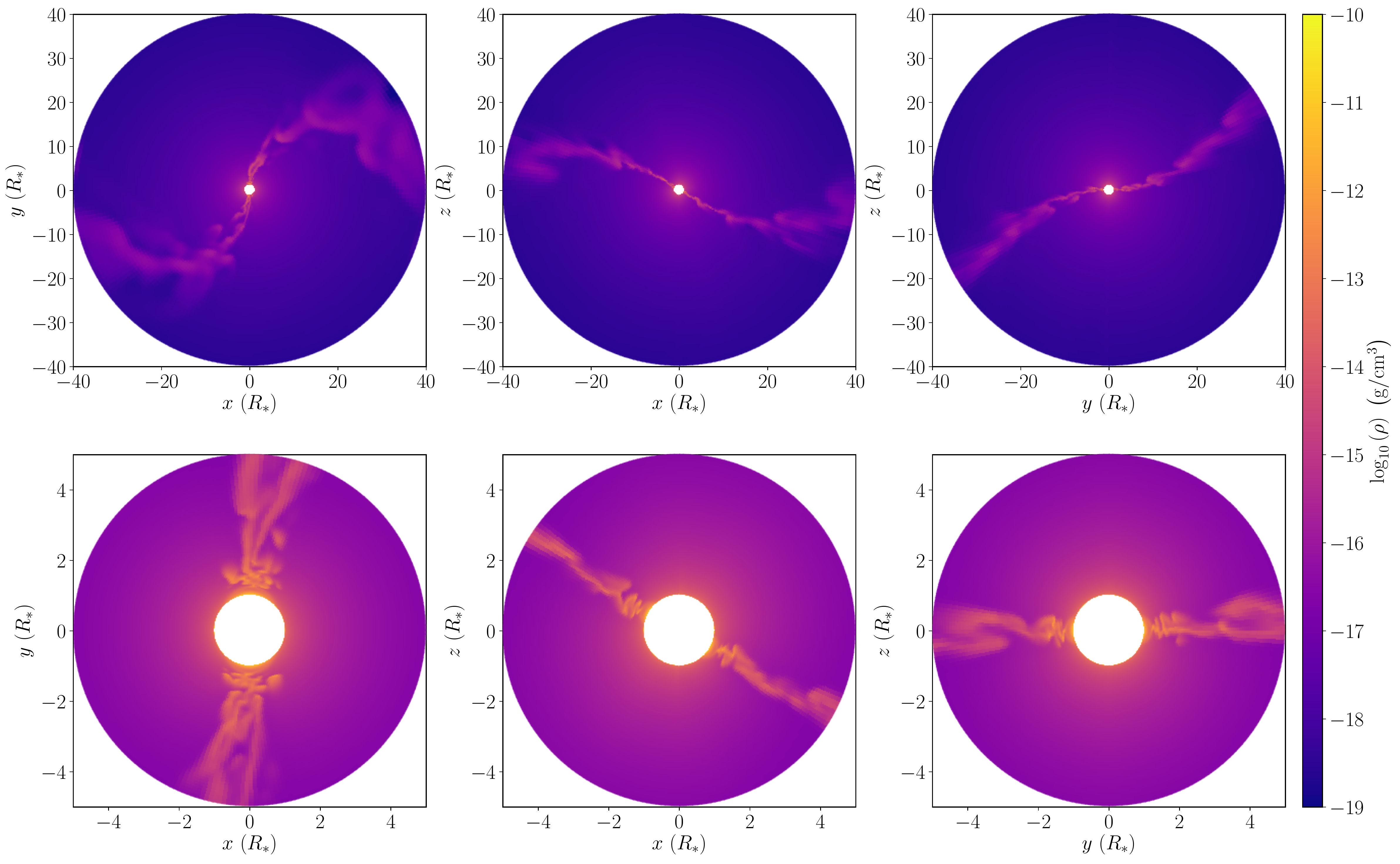}
	\caption[Density slice plots of the quasi-steady state density structure at $1 \ \mathrm{Ms}$.]{Slice plots of the global quasi-steady state density structure at $1 \ \mathrm{Ms}$. The star is situated in the centre with the confined material appearing as a curved "S" shape. This is due to the disk-like structure intersecting the plain of the slice. Each plot shows a slice in a different coordinate plane, from left to right is the; $xy$-, $xz$- and $yz$-plane. All three columns exhibit the contortion of the expanding excretion disk due to the action of rotation. Top: the full simulation domain. Bottom: close-up of the density structure of the inner $5 \ R_{\ast}$ of the simulation. The density structure clearly shows the confinement of material in the magnetic equator, off-set from the rotational equator.
		\label{fig:density_multi}}
\end{figure*}

\begin{figure*}
	\centering
	\includegraphics[width=1.0\textwidth]{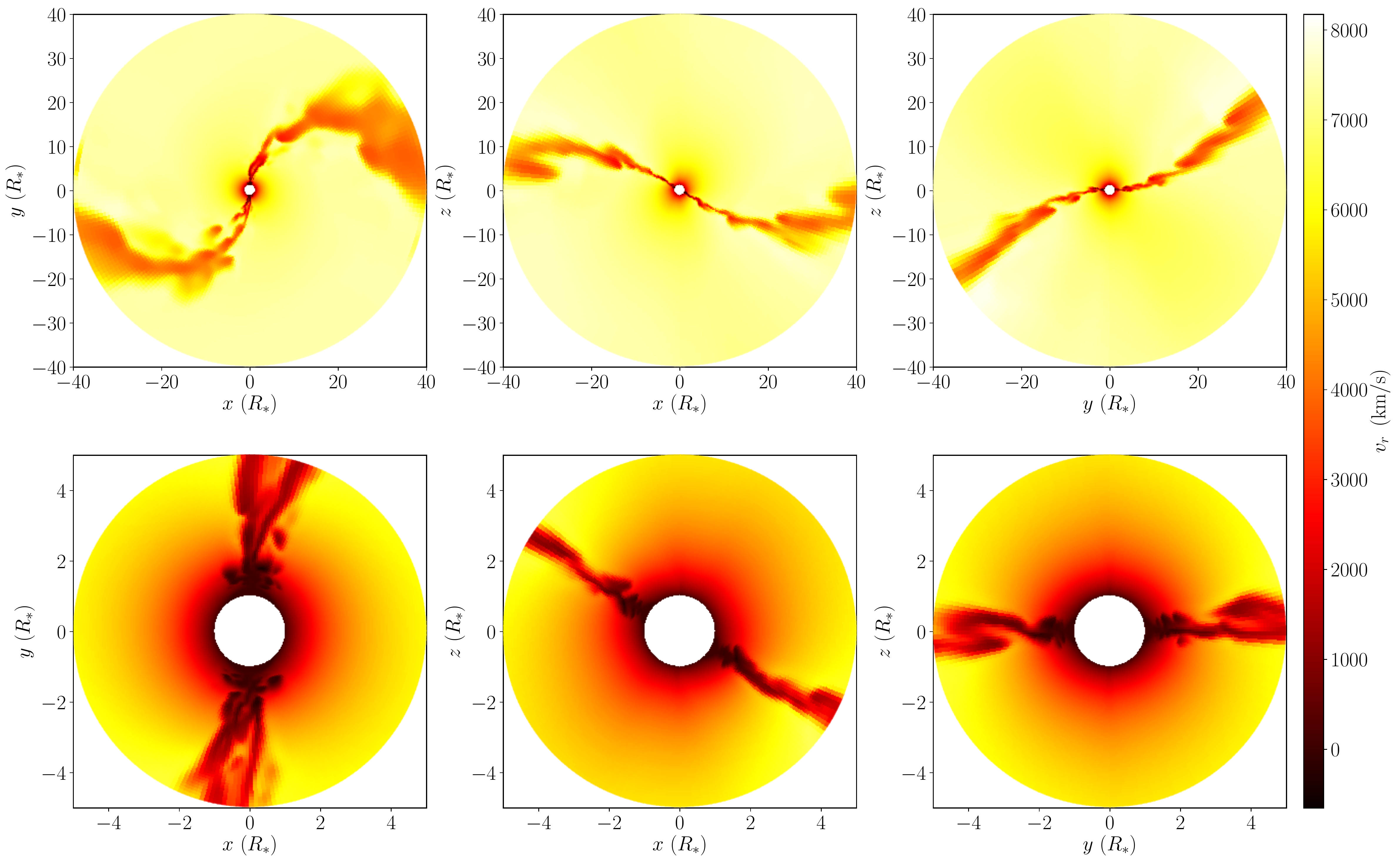}
	\caption[Velocity slice plots of the quasi-steady state density structure at $1 \ \mathrm{Ms}$.]{Same as Fig. \ref{fig:density_multi} but for the velocity magnitude profile. The top row shows the extended wind with a clear contrast between the free streaming and the slower moving disk material. The bottom row shows the inner wind velocity. The sonic point is reached within a fraction of a stellar radus. The polar axis is visible as a discontinuity in values in the central panel of the bottom row. This jump has however had a negligible impact on the extended wind, as can be see in the cosponsoring panel of the top row. Beyond $\sim5 \ R_{\ast}$, the wind has reached the terminal velocity. The profile in all three panel exhibits a value approximatly twice that expected for the terminal velocity, see section \ref{sec:comparison} for discussion of this.
		\label{fig:velocity_multi}}
\end{figure*}

\begin{figure*}
	\centering
	\includegraphics[width=1.0\textwidth]{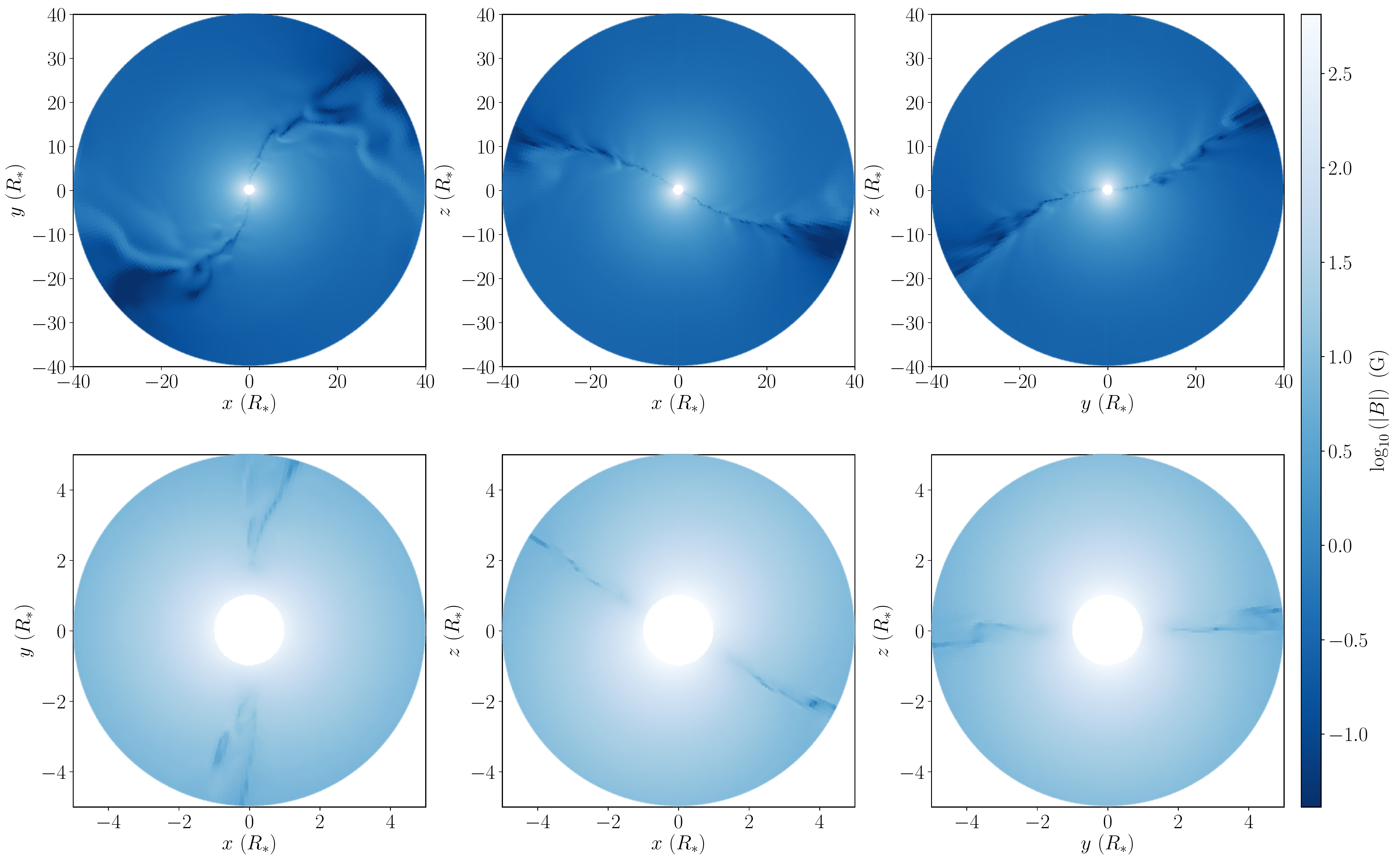}
	\caption[Mangetic slice plots of the quasi-steady state density structure at $1 \ \mathrm{Ms}$.]{Same as Fig. \ref{fig:density_multi} but for the $\log_{10} (|\bm{B}|)$. The central column shows a slice looking down the $y$-axis which is the axis about which the dipole field is rotated; the maximum field magnitude is located on the stellar surface at both magnetic poles. This value is $648 \ \mathrm{G}$, twice the equatorial value. Examining the top row, it can be seen that the magnetic field decays smoothly from the surface to the extended wind everywhere except for the excretion disk, where the current sheet is and is several orders of magnitude lower than in the free streaming wind.
		\label{fig:B_field_multi}}
\end{figure*}

3D representations of the star and inner magnetosphere showing the stellar surface, magnetic field lines and isodensity surface are given in Fig. \ref{fig:volume}. Two arrows indicate the rotation axis (vertical arrow) and the magnetic axis (oblique arrow). The yellow isodensity contour illustrates the confinement of material in the closed field region.

The radial extent of this region and the closure latitude are related through the following equation:
\begin{equation}
\sin (\theta_{\mathrm{c}}) = \sqrt{R_{\ast}/R_{\mathrm{c}}}.
\label{eq:closed_field}
\end{equation}
Where $R_{\mathrm{c}}$ is the radial distance from the centre of the star to the apex of the longest closed field line and $\theta_{\mathrm{c}}$ the co-latitudinal coordinate at which the field line makes contact with the stellar surface. The subscript $_{c}$ stands for closure \citep{ud-Doula2008}. \cite{Vidotto2011} use a similar expression to estimate the latitude of cyclotron emission from an exoplanetary atmosphere using the notation $R_{\mathrm{M}}$ indicating the radius of the magnetosphere. As massive stars are the topic of the present study we use the notation of \cite{ud-Doula2008}. As an estimate for $R_{\mathrm{c}}$, we use the mean Alfv\'{e}n radius calculated from our simulation results, $R_{\mathrm{A}} = 3.98 \ R_{\ast}$, which leads to a closure latitude $\theta_{\mathrm{c}}~=~\pm 30.1^{\circ}$. This analysis neglects deformation of the closed field lines by either rotation or wind ram-pressure. By visual inspection of Fig. \ref{fig:volume}, there is qualitative agreement between the prediction of  $R_{\mathrm{c}}~\sim~3 \ R_{\ast}$ and $\theta_{\mathrm{c}}~\sim~\pm30^{\circ}$ and our simulation results.

Another feature shown in Fig. \ref{fig:volume} is the break of symmetry about the rotational axis due to the obliquity of the dipole. \cite{ud-Doula2013} observed symmetry breaking in the excretion disk, however as their simulation included no dipole obliquity, this breakdown of symmetry is due to the interplay between rotation and optically thin radiative losses. This final effect is not present in the current study and any rotational asymmetry is therefore due to the magnetic field topology.

Fig. \ref{fig:density_multi}, \ref{fig:velocity_multi} and \ref{fig:B_field_multi} show multiple slices through the computational domain in the $xy$-, $xz$- and $yz$-planes for $\rho$, $|\bm{v}|$ and $|\bm{B}|$ respectively for both the extended and inner wind regions. In each plot the star is centred at the origin and the wind extends from the surface at $r~=~1 \ R_{\ast}$ to $r~=~40 \ R_{\ast}$ where it leaves the simulation domain.

\subsubsection{Density}

The 2D density profiles in Fig. \ref{fig:density_multi} further emphasis the sharp departure from both the initial spherical wind and the cylindrical symmetry seen in the aligned dipole case. In the central column, in which we look down the $y$-axis, the magnetic field obliquity is clearly visible as the arms of the excretion disk are off-set from the equatorial plane by $30^{\circ}$, the same off-set as the magnetic field dipole vector. The panel on the left looks down the rotational axis and shows a slice along the rotational equator. As the magnetic equator is off-set from this, the slice cuts through the excretion disk, which appears as a contorted S shape. All panels clearly show the confinement of material in the magnetic equator, off-set from the rotational equator, which expands radially to form an extended excretion disk. As the rotation of the star processes, this disk is then contorted.

\subsubsection{Velocity field}
\label{sec:velocity_profile}

For the velocity profile in Fig. \ref{fig:velocity_multi}, there is a clear contrast between the free streaming and the slower moving disk material with a difference in velocity magnitude of the order $\sim~2000 \ \mathrm{km/s}$. However, the entire simulation is supersonic, with the sonic point virtually indistinguishable from the stellar surface.

A faint but important numerical feature, visible in the central column on the velocity profiles, is a discontinuity across the polar axes. Both plots show a non-physical jump in values due to the latitudinal boundary conditions. However, this discontinuity in the velocity has not propagated into the extended wind and therefore we assume there is negligible impact on the wind evolution.

Preliminary adiabatic simulations showed that this polar axis discontinuity results in spurious heating along the pole. As the simulation evolved, this numerical thermal perturbation begins to impact all fluid quantities. This is the primary reason for choosing  an isothermal model in which the energy and therefore temperature is constant.

The velocity profile, according to the prediction of CAK theory, should reach a terminal velocity of approximately $v_{\infty}~=~3000 \ \mathrm{km/s}$. This is not however what is observed in the simulation; with the extended velocity profile reaching $v_{\infty}~\approx~8000 \ \mathrm{km/s}$. This is approximately thrice that predicted by CAK theory, this feature has been discussed at length in Section \ref{sec:comparison}. The isothermal nature of the model imposes several restrictions on the physics involved in the radiative driving. Line acceleration is quenched in regions where the gas temperature exceeds $\sim~10^{6} \ \mathrm{K}$ as the line transitions in which the stellar flux is scattered are no longer available due to virtually full ionisation \citep{Pittard2009}. Temperatures of this magnitude have been modeled in 3D simulation of magnetic O-star winds, where gas is shock heated in the disk \cite{ud-Doula2013}. The isothermal condition used in our simulation prevents this shock heating and the temperature is restricted to the stellar effective temperature, well below the ionisation cut-off. Wind speeds seen in our simulations maybe due to this limitation. However, as the radio/sub-mm emission is primarily a function of the density field, we conclude that the velocity fields departure from what is expected has a negligible impact on the results.

\subsubsection{Magnetic field}

The final set of profiles, displayed in Fig. \ref{fig:B_field_multi}, depict the magnetic field in the three axes planes. The central plane clearly shows the obliquity of the dipole with the largest values indicating the two magnetic poles, off set from the rotational poles by $\zeta~=~30^{\circ}$. This magnetic polar field is $648 \ \mathrm{G}$, twice the equatorial value. The magnetic field decays smoothly from the surface to the extended wind everywhere except for the excretion disk, where the current sheet has formed. Within the Alfv\'{e}n radius, close to the star, the magnetic field controls the flow dynamics, however for the vast majority of the extended wind the ram pressure, $\rho v^{2}$, dominates.

\subsection{Spherical nature of the wind}
\label{sec:sph_wind}

\begin{figure}
	\centering
	\includegraphics[width=0.49\textwidth,trim={0cm 1.5cm 0cm 2cm},clip]{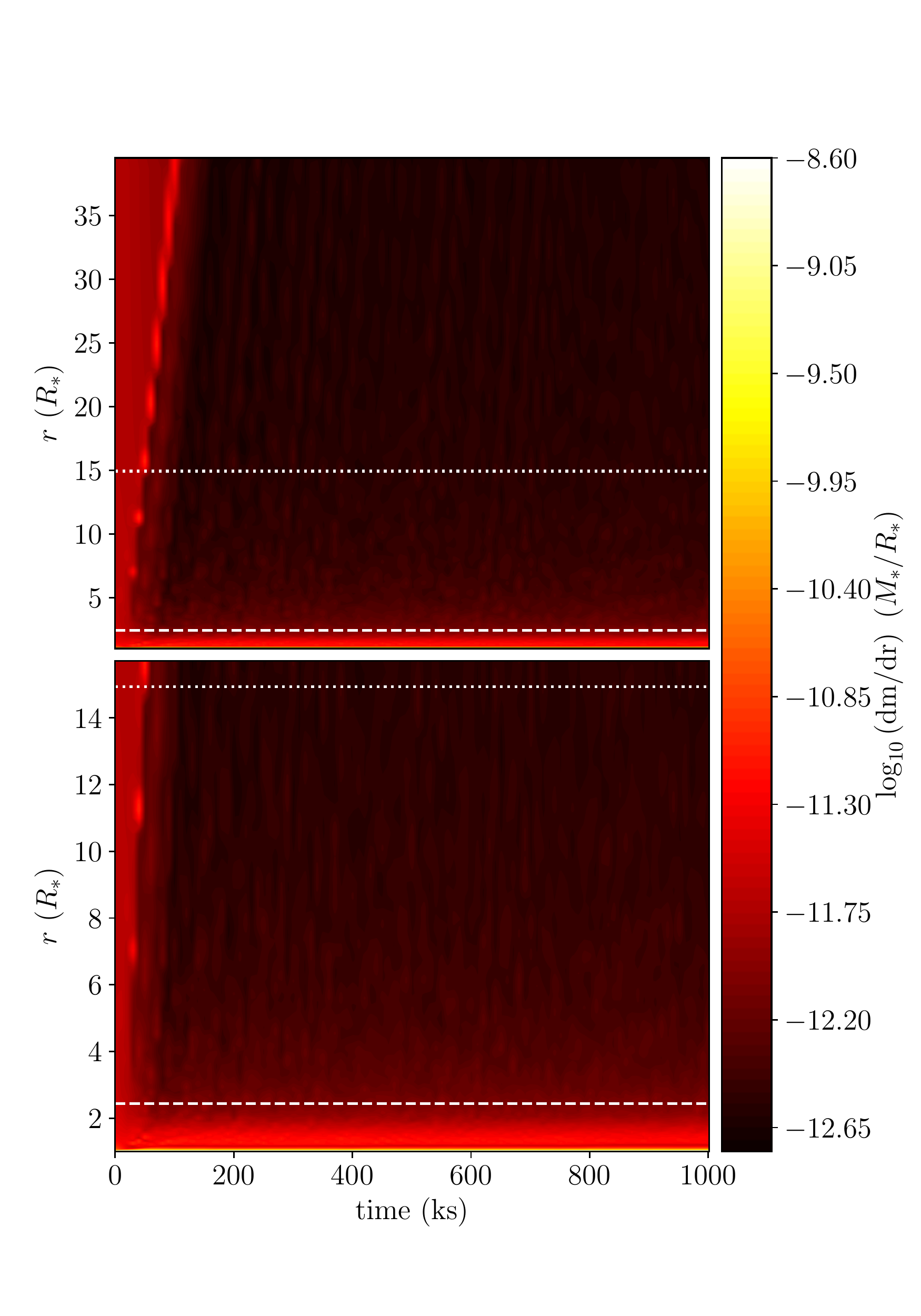}
	\caption[Radial mass distribution for the total simulation and the inner $16 \ R_{\ast}$]{Radial mass distribution for the total simulation (top panel) and the inner $16 \ R_{\ast}$ (bottom panel). The blow-off of the initial conditions can be seen in the left of the top panel, where a fan of material tracks outwards in the first $~\sim~200 \ \mathrm{s}$. Beyond this point the global radial motion of material is approximately constant with small perturbations resulting in lines tracking outward from the stellar surface coursed by clumps which form in the inner magnetosphere. In both panels the Alfv\'{e}n (dashed line) and Kepler (dotted line) radii are marked.
		\label{fig:dmdr}}
\end{figure}

\begin{figure}
	\centering
	\includegraphics[width=0.49\textwidth,trim={0cm 0.5cm 0cm 0cm},clip]{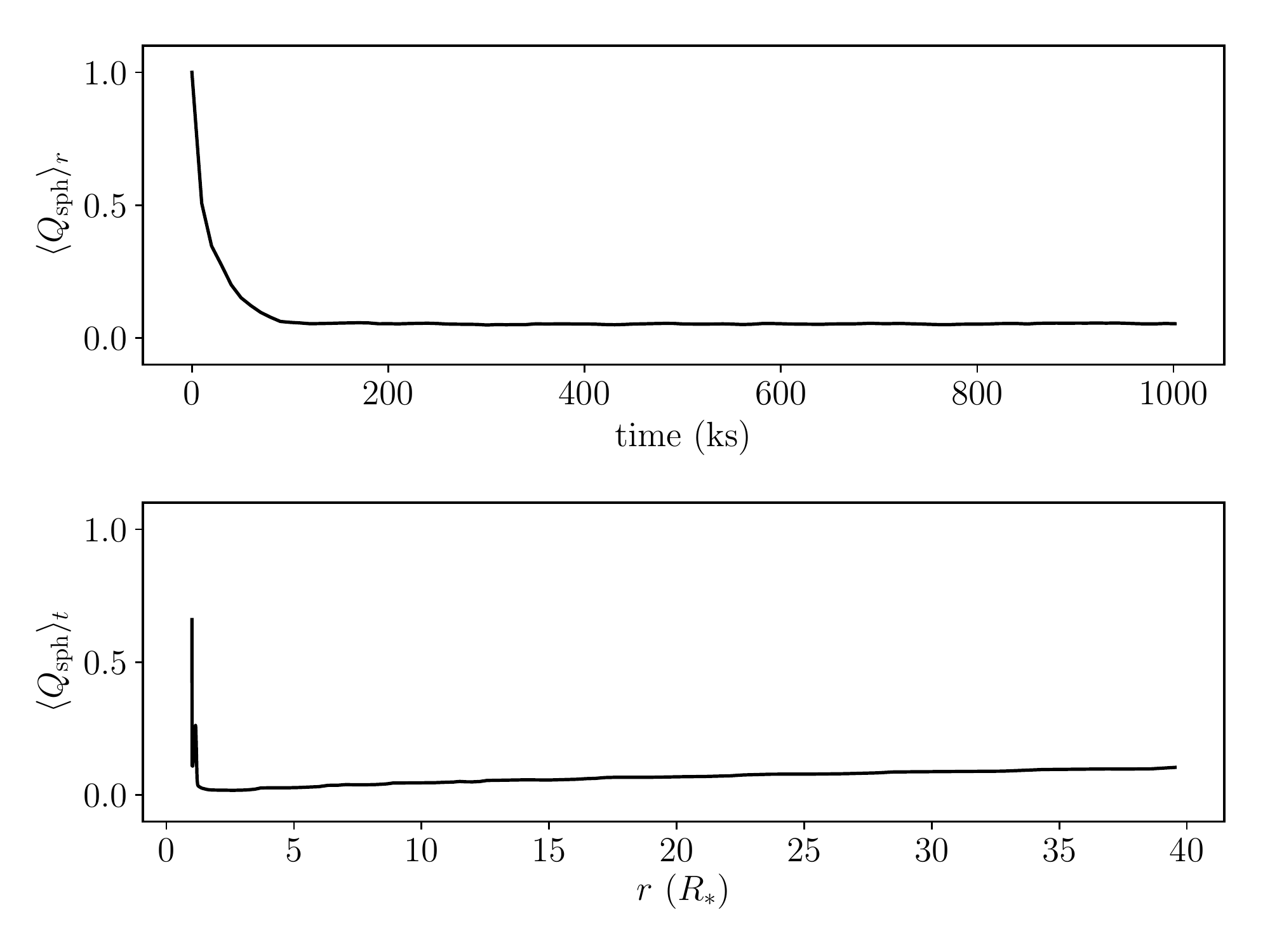}
	\caption[Average spherical quality factor, $\langle Q_{\mathrm{sph}} \rangle$, averaged over both time and space.]{Average spherical quality factor, $\langle Q_{\mathrm{sph}} \rangle$, averaged over both time (top) and space (bottom). Both plots show a rapid departure from spherical symmetry within the first $100 \ \mathrm{s}$. While the time average falls to 0.055 and remains approximately constant, the spatial average initially decreases then linearly increases out to $40 \ R_{\ast}$ where it reaches a value of 0.104, meaning that the inner wind is less spherical than the outer.
		\label{fig:Qsp_av}}
\end{figure}

At this point we shall discuss the extent to which the wind has deviated from its initial spherical symmetry.

As the calculation of thermal radio emission is a function of the maximum optical depth along the line of sight and this in turn is a function of the column density, the wind density structure entirely determines both the magnitude and rotational modulation of the observed spectral flux. Wind clumping, collimation and disk structures will all impact the degree to which emission can escape the system and reach the observer.

In the case of a purely spherical wind, there is an effective minimum radius around the star at which emission from the wind at larger radii can escape to the observer; emission from within the effective radius cannot. The optical depths dependence on the density, means that emission from the inner wind is effectively (from the point of view of the observer) obscured by the extended wind at larger radii. This effective radius is thus the radio photosphere of the star. For a clumped, magnetically confined or otherwise non-spherical wind, this photosphere is not spherical and emission escapes from varying radii. As such, quantifying the winds departure from spherical symmetry is an important step in placing the radio emission in context.

This is accomplished by following the approach of \cite{ud-Doula2008}, who devised an expression to quantify the radial distribution of material in the magnetosphere. By integrating the product $r^{2} \rho\left( r, \theta, \phi, t \right) \sin(\theta)$ over the two angular coordinates, $\theta$ and $\phi$, one is left with the global density structure stratified in the radial direction. This radial mass distribution is given by the following expression:
\begin{equation}
\frac{\mathrm{d} m \left( r, t \right)}{\mathrm{d} r} = r^{2} \int^{2 \pi}_{0} \int^{\pi}_{0} \rho \left( r, \theta, \phi, t \right) \sin(\theta) \mathrm{d} \theta \mathrm{d} \phi.
\label{eq:dmdr}
\end{equation}
Motivated by the need to capture the behaviour of material in the case of an aligned dipole, \cite{ud-Doula2008} chose to limit the integration over $\theta$ to a small angular region centred about the equator. In this study, the behaviour is not constrained to the rotational equator so we integrate over the full range of $\theta$.

Equation (\ref{eq:dmdr}) is plotted in Fig. \ref{fig:dmdr} over the entire radial range of the simulation and for a limited region of the inner $6 \ R_{\ast}$. The first $\sim~100 \ \mathrm{ks}$ of the evolution of $\mathrm{d} m \left( r, t \right) / \mathrm{d} r$ is a striking illustration of the blow-off of the initial conditions, where a fan of higher density material tracks outwards from the surface to the boundary. For $t~>~100 \ \mathrm{ks}$ the global radial motion of material is approximately constant with only small perturbations, as clumps of material concentrate in the closed magnetosphere, breakout and track outwards leading to radial lines in $\mathrm{d} m / \mathrm{d} r$.

Both the Kepler and Alfv\'{e}n surfaces are indicated in the figure and there is little change in $\mathrm{d} m  / \mathrm{d} r$ across either. This is consistent with the dynamical magnetospheric behaviour described in Section \ref{sec:magnetosphere}.

As a consequence of the magnetic confinement and the reduction in the mass-loss rate, the stellar wind is overall much less dense than for a corresponding non-magnetised stellar wind (this is consistent with Fig. \ref{fig:comparison}) and emission from deeper in the wind, closer to the stellar surface will be able to escape. However, as the total wind material is reduced, the total spectral flux, $S_{\nu}$, will also be reduced; resulting in a fainter signal reaching the observer. Additional free-free absorption along the line of sight may also contribute to the reduction in observable emission.

To further quantify the departure from spherical symmetry, we now detail the formalism of a dimensionless measure used to indicate the overall spherical nature of the wind.

For a given radius there exists a spherical shell, $S$, of width $\mathrm{d}r$. To quantify the spherical distribution of material within this shell we adopt the following procedure. Each density value within the shell, $\rho \left( \theta, \phi \right)$, is normalised by the maximum density, $\rho_{\rm{max}} \left( \theta, \phi \right)$, in the shell; these normalised density values are then summed over the spherical shell for all vales of $\theta$ and $\phi$. Finally this summation is divided by the total number of sample points within the shell giving the average normalised density in the shell. The final expression,
\begin{equation}
Q_{\mathrm{sph}} \left( r, t \right) = \frac{1}{N_{S}} \sum_{\theta} \sum_{\phi} \frac{\rho \left( \theta, \phi \right)}{\rho_{\mathrm{max}} \left( \theta, \phi \right)},
\label{eq:Q}
\end{equation}
gives the spherical quality factor, $Q_{\mathrm{sph}}$ (not to be confused with the Q-factor, $\overline{Q}$, in equation (\ref{eq:line_accel})), within the range $0~<~Q_{\mathrm{sph}}~<~1$ and is a measure of the departure of the density distribution from a spherical wind within shell $S$ at radius $r$ and time $t$. A value of 1 indicates a spherically symmetric wind while a value of 0 indicates a complete departure from spherical symmetry, a disk structure akin to a delta function. This is un-physical and in reality Q would be asymptotic to 0 but never reach it.

By computing $Q_{\mathrm{sph}}$ for every radial shell and for every time point, it is then possible to calculate both the time average, $\langle Q_{\mathrm{sph}} \rangle_{t}$, for every radial point and the radial averaged, $\langle Q_{\mathrm{sph}} \rangle_{r}$, for every time point. These two quantities are plotted in Fig. \ref{fig:Qsp_av}. Each data point in the top plot represents the $\langle Q_{\mathrm{sph}} \rangle_{r}$ for the entire simulation volume as a function of time and every data point in the bottom plot represents the $\langle Q_{\mathrm{sph}} \rangle_{t}$ across every time point in the simulation as a function of radius.

From Fig. \ref{fig:Qsp_av}, we can see that, for all time after the initial $\sim~100 \ \mathrm{ks}$, $\langle Q_{\mathrm{sph}} \rangle_{r}$ remains constant at $0.055$ (consistent with the results in Fig \ref{fig:comparison} and Fig. \ref{fig:dmdr}). However, the radial profile of the time average indicates an increase in $\langle Q_{\mathrm{sph}} \rangle_{t}$ towards larger radii to a value of $0.104$. This is consistent with a broadening out of the excretion disk as it expands radially.

Temperature is constant in the simulation, however as the disk is denser than its surroundings there is still a pressure gradient leading to an expansion, this together with rotation and diverging magnetic field lines spreads material out at larger radii. For an adiabatic model, this expansion will be made more acute as a temperature would also enhance the broadening and lead to a greater $\langle Q_{\mathrm{sph}} \rangle_{t}$ at larger radii than seen here.

The results for both $\mathrm{d} m \left( r, t \right) / \mathrm{d} r$ and $\langle Q_{\mathrm{sph}} \rangle$ indicate a wind which has undergone rapid departure from spherical symmetry, with a decrease in the overall wind mass from the initial condition leading to lowered column density along the observers view compared to the same wind without the action of a magnetic field. With this in mind, we now turn to the results of the synthetic radio emission.

\begin{figure*}
	\centering
	\includegraphics[width=1.0\textwidth]{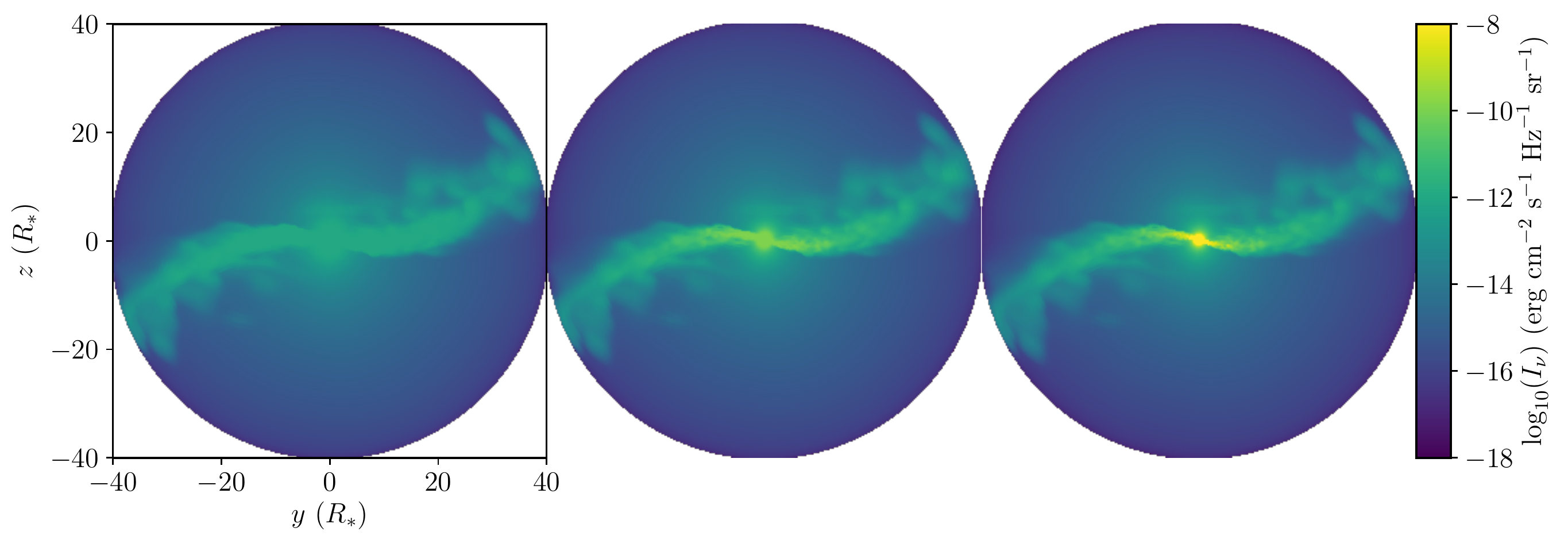}
	\caption[Intensity of radio emission for an observing inclination of $30^{\circ}$ and phase $216^{\circ}$ at increasingly higher frequency.]{Intensity of radio emission for an observing inclination of $30^{\circ}$, phase $216^{\circ}$ and observing frequencies of $10 \ \mathrm{GHz}$ (left), $100 \ \mathrm{GHz}$ (middle) and $1000 \ \mathrm{GHz}$ (right). The figure illustrates the additional resolution gained by observing at higher frequencies, as the panel, from left to right shows a progressively sharper central star.
		\label{fig:radio_wind_freq}}
\end{figure*}

\begin{figure*}
	\centering
	\includegraphics[width=0.665\textwidth]{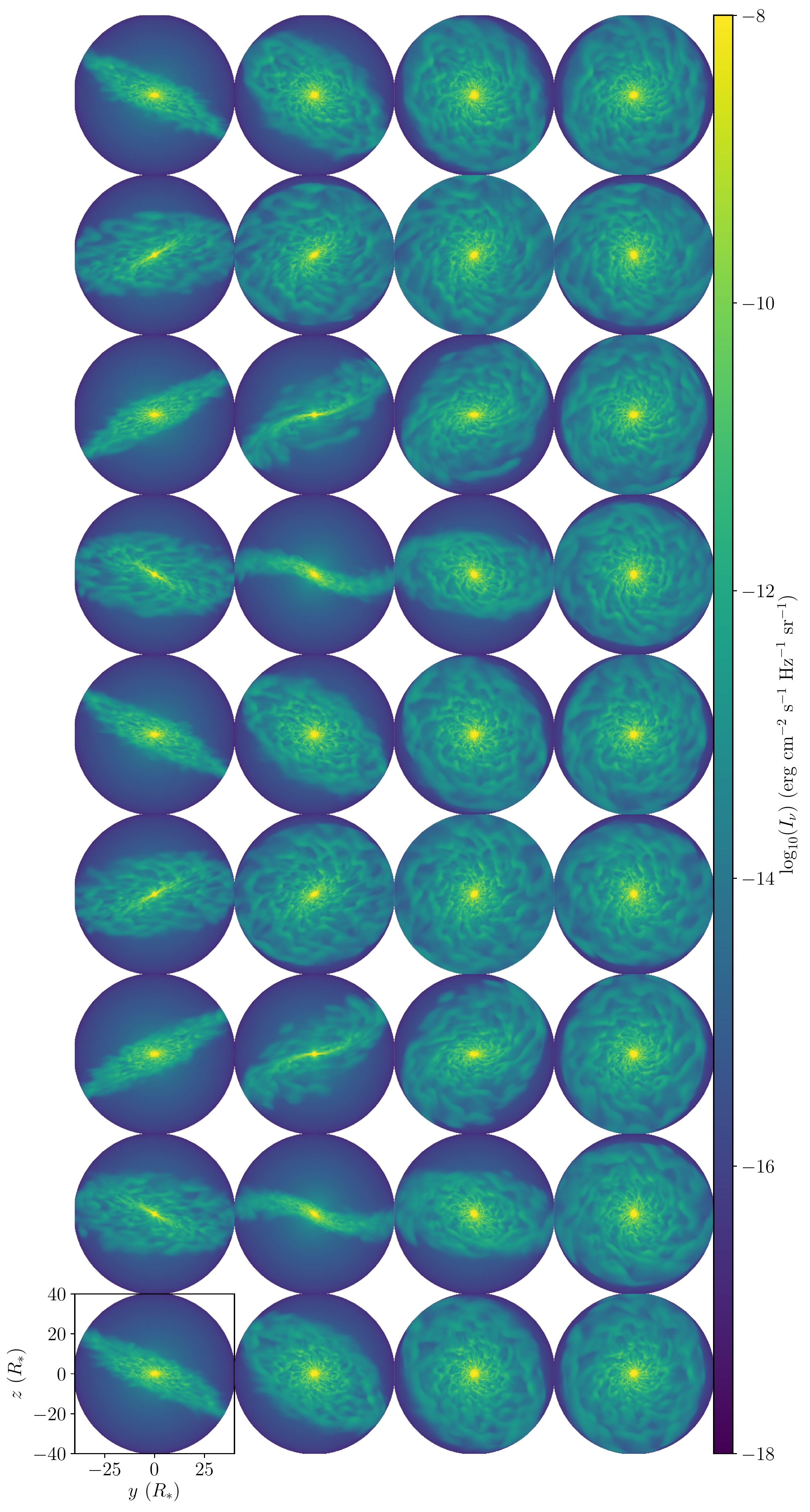}
	\caption[Intensity of radio emission for rotational phase and observing inclination at $900 \ \mathrm{GHz}$.]{Intensity of radio emission for rotational phase and observing inclination, all plots are generated at $900 \ \mathrm{GHz}$. Each column represents observing inclinations of $0^{\circ}$ (viewing along equator), $30^{\circ}$, $60^{\circ}$ and $90^{\circ}$ (viewing down the polar axes) from left to right respectively. Each row are snap shots of the radio intensity for increasing phase from $0 \ \mathrm{radians}$ to $4\pi \ \mathrm{radians}$ (two full rotations), equally spaced by $4\pi/9 \ \mathrm{radians}$. The second column on the left experiences the greatest degree of variability in the disk while the right most column experiences the least, presenting the same disk surface area to the observer over the 2 complete rotations.
		\label{fig:radio_wind_phase_incl}}
\end{figure*}

\subsection{Radio/sub-mm emission}
\label{sec:emission_res}

We divide the radio emission results into first a discussion of the emission volume structure as it appears from the radio intensity calculation of equation (\ref{eq:I_tau_max}), together with its dependence on observing frequency, inclination and rotational phase. Secondly, the total spectral flux density, $S_{\nu}$, is calculated to give radio light curves for two full stellar rotations, at three discreet frequencies over a range of phases and inclinations. We then calculate the continuum emission over the frequency range $10 \ \mathrm{GHz}~<~\nu~<~10^{4} \ \mathrm{GHz}$ at constant inclination and phase. This range is chosen as it spans the observing bands of the ALMA which cover the frequency range 84 - 950 GHz (Band 3 - Band 10) and beyond into infrared wavelengths. Currently this range is covered by SOFIA (Stratospheric Observatory for Infrared Astronomy), however its sensitivity maybe insufficient for all but the brightest O-stars.

For both the light curves and continuum spectra, we calculate ratio of the mass-loss derived from the synthetic radio calculations to that measured directly in the simulation, $\dot{M}_{\mathrm{obs}} / \dot{M}_{\mathrm{sim}}$, allowing us to compare what would be inferred, via radio observation, to the actual mass-loss rate of the simulated star.

\subsubsection{Wind structure in emission}
\label{emission_structure}

To illustrate the concept of optical depth and dependence of observational results on the chosen observing frequency, we plot in Fig. \ref{fig:radio_wind_freq} the intensity of radio emission, $I_{\nu}$, at a phase of $30^{\circ}$ and inclination of $216^{\circ}$ for three dex in observing frequency, $\nu$: $10 \ \mathrm{GHz}$, $100 \ \mathrm{GHz}$ and $1000 \ \mathrm{GHz}$. Form left to right, the figure shows increasing fidelity in the inner region close to the star, where the density profile has its largest values. Higher frequencies thus penetrate further into the wind. Only the highest observing frequency of 1000 GHz is able to penetrate the wind down to approximately the stellar surface (see \citealt{Daley-Yates2016} for a discussion on the frequency dependence of the radio photosphere radius). If $\nu$ is increased beyond $1000 GHz$, emission from the stellar surface black body radiation will begin to dominate, we will discuss this further in Section \ref{sec:continuum}

To gain an appreciation of the rotational modulation and dependence on observer inclination of the intensity map, $I_{\nu}$, we limit the parameter space to a single observing frequency of $900 \ \mathrm{GHz}$; chosen as it provides the highest fidelity images, probing deeper into the wind, while still within ALMA frequency band 10. Fig. \ref{fig:radio_wind_phase_incl} shows time series snapshots for $I_{\nu}$. Rows 1 - 9 show the rotational phase at equidistant intervals of $4\pi/9$ radians across the full two rotations and the four columns show. From left to right, inclination values of $0^{\circ}$, $30^{\circ}$, $60^{\circ}$ and $90^{\circ}$.

The second column, inclination of $30^{\circ}$, shows the greatest degree of variability with the disk presenting both its edge and face to the observer. Looking down the rotational axis, inclination of $90^{\circ}$, the disk exhibits the smallest degree of variability, always showing the same disk extent. However, a face on disk presents the largest ratio of visible surface area to volume for an observer; resulting in the greatest total emission.

For all phases and inclinations, the central star is clearly visible, in contrast to the left image of Fig. \ref{fig:radio_wind_freq}, calculated at $10 \ \mathrm{GHz}$, where the star is obscured by the extended wind and emission from the centre is of the same order of the surrounding disk. As Fig. \ref{fig:radio_wind_phase_incl} is calculated at $900 \ \mathrm{GHz}$, we can conclude that this frequency is sufficient to probe the range of densities, and therefore optical depths, occurring in the simulation.

A final noteworthy result of the rotation and inclination calculation, is the apparent difference of the rotational period between the first column of Fig. \ref{fig:radio_wind_phase_incl} and the other three. For this first column, rows 1, 5 and 9 show the same image, while rows 3 and 7 show the same image but inverted about the $z$-axis. This inversion is evident form the simulation results, as the intensity map effectively shows a resolved source. However, an earth bound observer sees the total flux, $S_{\nu}$, which is the integration of $I_{\nu}$ over both $y$ and $z$. Such inversions of the source are not captured by the total flux and may lead to false predictions of the rotational period, as shown in the following section.

\subsubsection{Radio lightcurves}
\label{sec:rotation}

\begin{figure*}
	\centering
	\includegraphics[width=0.85\textwidth]{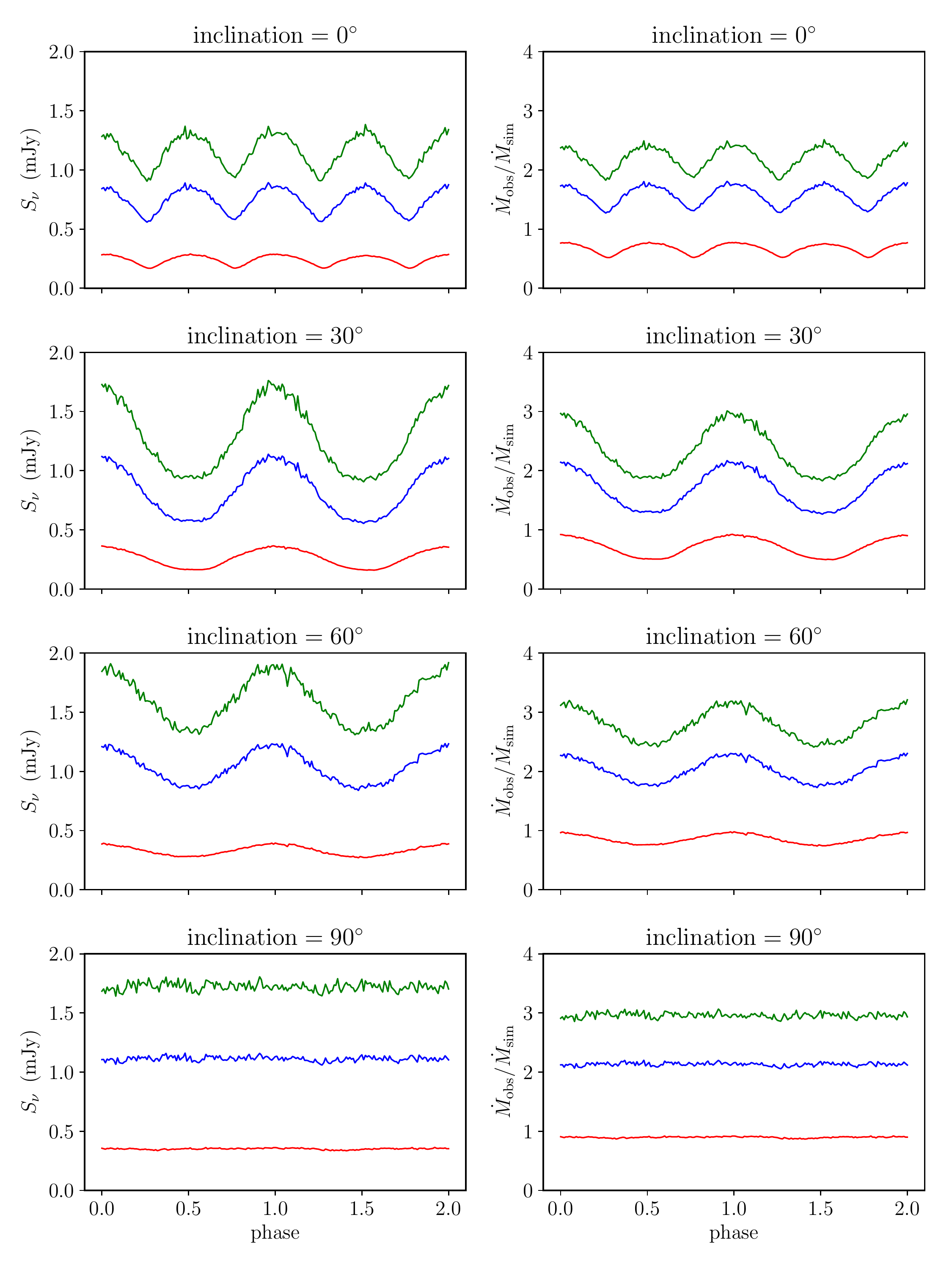}
	\caption[Radio light curves and recovered mass-loss rate for the simulated O-star and varying frequencies and inclination.]{Left column: radio lightcurves over two rotations for the 4 observing inclinations, top to bottom: $0^{\circ}$ (viewing along equator), $30^{\circ}$, $60^{\circ}$ and $90^{\circ}$ (viewing down the polar axes). Each inclination shows the result for the 3 observing frequencies; $250 \ \mathrm{GHz}$ (red), $650 \ \mathrm{GHz}$ (blue) and $900 \ \mathrm{GHz}$ (green). What is immediately apparent is the lack of variability for all three frequencies in the $90^{\circ}$ inclination plot. This is due to the excretion disk presenting the same surface area to the observer over the 2 complete rotations. For an inclination of $0^{\circ}$, the light curve exhibits 4 distinct and maxima, in contrast to $30^{\circ}$ and $60^{\circ}$ inclinations which only show 2 distinct minima and maxima. Right column: rotational modulation of the inferred mass-loss from the synthetic radio lightcurves. All frequencies and observing inclinations show modulation of the predicted mass-loss except observations at $90^{\circ}$ which shows flat predictions at all phases and frequencies. Rotation at $900 \ \mathrm{GHz}$ and inclination of $30^{\circ}$ shows the largest degree of modulation. Each observing frequency results in a different inferred mass-loss due to the dependence of mass-loss on the spectral flux.
		\label{fig:rotation}}
\end{figure*}

\begin{figure*}
	\centering
	\includegraphics[width=1.0\textwidth]{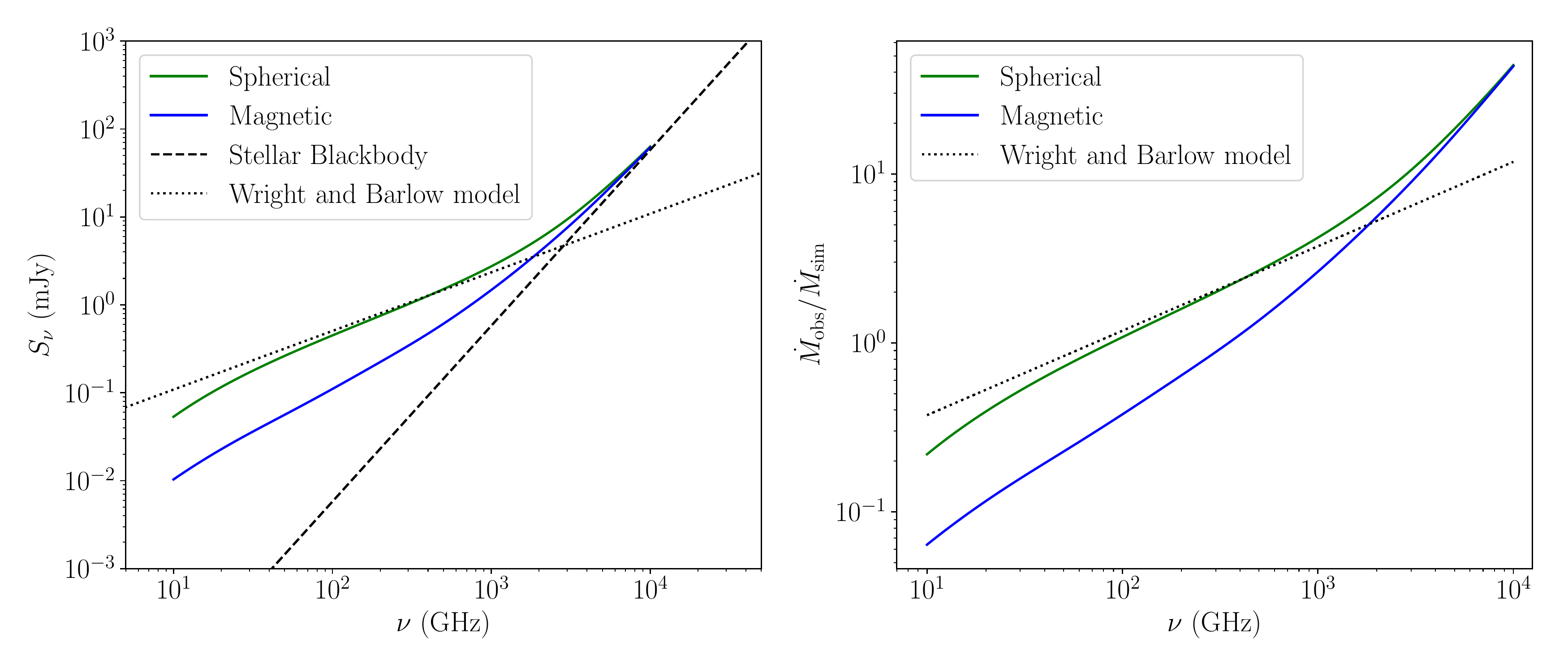}
	\caption[Radio contimum spectra and recovered mass-loss for the simulated O-star.]{Left: radio/sub-mm spectrum for 3 models of emission; numerical spherically symmetric (green curve), numerical magnetic (blue curve) and analytic WB75 model (black dotted line). The stellar surface black-body curve is also shown (black dashed line). Both numerical models converge with the surface black-body at $\nu > 10^{4} \ \mathrm{GHz}$ while the WB75 model does not. Right: Corresponding mass-loss predictions for the three emission models using equation (\ref{eq:WB_M_dot}).
		\label{fig:spectrum}}
\end{figure*}

We now turn to the time dependence of $S_{\nu}$. Fig. \ref{fig:rotation} shows this dependency for the rotational phases and inclination of Fig. \ref{fig:radio_wind_phase_incl} for observing frequencies of $250 \ \mathrm{GHz}$ (Band 6) and $650 \ \mathrm{GHz}$ (Band 9) in addition to $900 \ \mathrm{GHz}$ (Band 10).

As already apparent in the intensity maps, $I_{\nu} (y, z)$, of the previous section; for an inclination of $90^{\circ}$ there is virtually no variability. This is in marked contrast to the curves of the three other inclination which all show rotational modulation between factors of approximately $1.5$ and $2$.

One would expect the greatest degree of modulation for the $30^{\circ}$ inclination, as this equals the obliquity of the dipole field and therefore the normal to the disk should coincide with the observers line of sight periodically through the rotation. This is indeed what we see in Fig. \ref{fig:rotation}. However, it is the $60^{\circ}$ inclination which results in the largest maximum emission. this is the result of a combination of magnetic tension and disk warping leading to the largest observed surface area for this inclination, at phases 0, 1 and 2.

For an inclination of $0^{\circ}$, the light curve exhibits 2 distinct minima and maxima per rotational cycle (a total of 4 distinct minima and maxima are present in the light curve as two rotational cycles are shown). This is in contrast to $30^{\circ}$ and $60^{\circ}$ inclinations which only show 1 distinct minima and maxima per rotational cycle (2 in total over the hole light curve of 2 rotations). Therefore, the top row does indeed shows an apparent rotation rate for the star which is twice the actual (if we take the time between peaks as the apparent rotation period), as predicted in the previous section.


The different behaviour of all lightcurves across all inclination and phases illustrates the sensitive dependence upon the magnetic field that the radio emission from the inner and extended wind has. Understanding observing inclination, rotational phase and obliquity is therefore critical for placing observed $S_{\nu}$ in context.

In Section \ref{sec:theory_radio} we detailed the method for estimating stellar wind mass-loss rates from the observed spectral flux, $S_{\nu}$. The right column of Fig. \ref{fig:rotation} shows the mass-loss rate inferred from the synthetic radio observations via equation (\ref{eq:WB_M_dot}), $\dot{M}_{\mathrm{obs}}$, normalised to the average simulation mass-loss rate, $\dot{M}_{\mathrm{sim}}$. As $\dot{M}_{\mathrm{sim}}$ is measured directly in our simulation, the resulting reduction in the wind mass and spherical nature due to magnetic confinement should be apparent and allow us to self-consistently assess the discrepancy between $\dot{M}_{\mathrm{sim}}$ and the inferred mass-loss from the synthetic radio emission, $\dot{M}_{\mathrm{obs}}$. Thus $\dot{M}_{\mathrm{obs}}/\dot{M}_{\mathrm{sim}} = 1$ corresponds to the situation where the inferred observed mass-loss from the synthetic radio emission is equal to the mass-loss measured directly in the simulation.

All frequencies and observing inclinations show modulation of $\dot{M}_{\mathrm{obs}}$ except for the $90^{\circ}$ inclination which shows flat predictions at all phases and frequencies consistent with the flat radio curves. These flat predictions still depart from $\dot{M}_{\mathrm{sim}}$ however, with higher frequencies overestimating the mass-loss rate. This becomes more acute as the observing frequency increases, with $900 \ \mathrm{GHz}$ leading to the largest overestimate. Rotation at $900 \ \mathrm{GHz}$ and inclination of $30^{\circ}$ shows the largest degree of modulation, constant with the corresponding $S_{\nu}$. As the mass-loss calculation, equation (\ref{eq:WB_M_dot}), has a functional dependence on the spectral flux of $\dot{M}_{\ast}~\propto~S_{\nu}^{3/4}$, we can expect for a doubling of $S_{\nu}$ to result in a $1.68$ increase in $\dot{M}_{\ast}$, which is approximately what we see for the lightcurves in the left-hand column of Fig \ref{fig:rotation}.

For the signature of magnetic confinement on the rotational modulation and therefore the radio emission and predicted mass-loss to diminish, the wind would have to return to a spherical expansion. For our simulation, this will happen at large radii (and therefore at low observing frequencies). From the lower plot of Fig. \ref{fig:Qsp_av}, we can see that $\langle Q_{\mathrm{sph}} \rangle_{t}$ has increased to 0.1 between the initial confinement, close to the star, and the outer simulation boundary. The physical distance required for $\langle Q_{\mathrm{sph}} \rangle_{t} \rightarrow 1$ is not covered in this work, however, it seems likely that this would happen at radii where the contribution to the spectral flux occurs at frequencies $<< 1 \mathrm{GHz}$. Once $\langle Q_{\mathrm{sph}} \rangle_{t}~=~1$, modulation of the radio emission would no longer happen. However, at this radii, the density of the wind would be so low that any contribution to the spectral flux may not be detectable. Additionally, the wind density would be insufficient to prevent rotationally modulated emission from closer to the star escaping to the observer. At these radii, the majority of the wind gas may have recombined and no longer be emitting, however as we do not calculate the radii at which $\langle Q_{\mathrm{sph}} \rangle_{t}~=~1$, we can not assert this and can only state that it will occur at radii much greater an that studied here ($40 \ R_{\ast}$). From these arguments we can conclude that, for the star simulated at the observing frequencies studied here, the extended wind of the star can not prevent emission escaping the inner wind and reaching an observer.

The noise seen in both the radio lightcurves and the corresponding mass-loss curves, is due to the numerical details of the radio calculation, which involves interpolation from spherical to Cartesian coordinate systems. By doing so, resolution is reduced close to the stellar surface where high frequency ($\sim~10^{3} \ \mathrm{GHz}$) emission originates from. Lower frequency ($\sim~10 \ \mathrm{GHz}$) emission originates from the outer wind where resolution is not reduced (for a spherical grid, the cell size increases with radius). This explains the lack of numerical noise in the $250 \ \mathrm{GHz}$ curves.

\subsubsection{Continuum spectrum}
\label{sec:continuum}

The radio continuum for our simulated star, along with two comparative models, is plotted in the left panel of Fig. \ref{fig:spectrum}. These comparative models are the WB75 model (black dots) discussed in Section \ref{sec:theory_radio} as well as the numerical results of the theory laid out in Section \ref{sec:theory_radio}, applied to the non-magnetic HD simulation (green curve). The stellar black-body radiation is also indicated (black dashes). The MHD wind result shows a clear departure from both the HD and WB75 models for $\nu~<~10^{3} \ \mathrm{GHz}$. At $\nu~>~10^{3} \ \mathrm{GHz}$, both the HD and the MHD curves converge to the black body radiation indicative of the optically thin regime of the stellar surface. This limit is not observed by the WB75 model as its theoretical basis ignores the presence of the stellar surface and the acceleration region where the density profile departs from a $1/r^{2}$ dependence (see \cite{Daley-Yates2016} for an in-depth discussion). This highlights the WB75 models applicability to the extended wind region, where a spherical wind will have a flat velocity profile (equal to $v_{\infty}$) and also its inability to capture the emission behavior at high frequency. The authors clearly stated the limitations of their model at high frequency.

Using order of magnitude arguments, we can determine the approximate limiting frequency of the WB75 model. This can be done by calculating a characteristic radio photosphere, introduced in Section \ref{sec:sph_wind}, which represents the minimum distance from the stellar surface which emission can escape from. The following analysis applies to a spherically symmetric wind only. For an in depth discussion of the effective radius see the original WB75 paper and \cite{Daley-Yates2016} for a more recent account. The effective radius, $R_{\mathrm{eff}}$, is calculated using equation (11) of WB75 and the stellar parameters used in this study as input (see Table \ref{tab:parameters}). $R_{\mathrm{eff}}$ was calculated for three observing frequencies, 1 GHZ, 10 GHz and 100 GHz with the results summarised in Table \ref{tab:R_eff}.

\begin{table}
	\caption{Comparison of characteristic radius for the effective radio-photosphere. \label{tab:R_eff}}
	\centering
	\begin{tabular}{ccc}
		\hline
		Observing frequency & $R_{\mathrm{eff}}$ $[R_{\sun}]$ & $R_{\mathrm{eff}}$ $[R_{\ast}]$ \\
		\hline
	    1 GHz & 317 & 35 \\
        10 GHz & 63 & 7 \\
        100 GHz & 13 & 1.4 \\
		\hline
\end{tabular}
\end{table}

The outer boundary of the simulation lies at 40 $R_{\ast}$, slightly larger than the effective radius of the star at 1 GHz. This means that activity from the inner magnetosphere is effectively obscured by the extended wind. Only when the star is observed at frequencies above 10 GHz does emission from the inner magnetosphere become appreciable. This is born out in Fig. \ref{fig:spectrum} where we see significant deviation from the WB75 model at observing frequencies $> 100 GHz$. The WB75 model is the basis for the mass-loss prediction of equation (\ref{eq:WB_M_dot}), its limitations will therefore effect any mass-loss predictions based on the spectral flux via this expression.

The corresponding normalised mass-loss prediction $\dot{M}_{\mathrm{obs}}/\dot{M}_{\mathrm{sim}}$ for the continuum spectra are displayed in the right panel of Fig. \ref{fig:spectrum}. All models show a dependence on $\nu$ with a two orders of magnitude variance. The HD model agrees with the WB75 model in the mid frequency range $10^{2} \ \mathrm{GHz}~<~\nu~<~10^{4} \ \mathrm{GHz}$, which corresponds to a constant gradient for the spectral flux (a spectral index of $\alpha~=~0.6$), but departs either side of this range. For the MHD model, except for agreement with the HD model at high frequency $\nu~>~10^{3} \ \mathrm{GHz}$, there is approximately an order of magnitude reduction in $\dot{M}_{\mathrm{obs}}/\dot{M}_{\mathrm{sim}}$ across all frequencies.

Recently \cite{Ramiaramanantsoa2017} communicated observations of the O4I(n)fp star $\zeta$ Puppis with the BRITE-Constellation nanosatellies. They detected one single periodic, non-sinusoidal component of the emission, which they attributed to the presence of bright surface features. Separate simultaneous spectroscopic observations led them to infer the action of corotating interaction regions (CIRs). While no surface spots are present in our simulation, the modelling of $\zeta$ Puppis by \cite{Ramiaramanantsoa2017} to explain the BRITE-Constellation observations, results in spiral structures of a similar nature to those in Fig. \ref{fig:density_multi}. We do not make direct comparisons to the models of \cite{Ramiaramanantsoa2017}, since $\zeta$ Puppis and our model star are very different. However, we highlight the similarity in the features and that, following further analysis of the synthetic observables at BRITE-Constellation frequencies, magnetically activity may provide a compelling explanation for the origin of CIRs in the magnetic massive star population.

JVLA observations by \cite{Kurapati2016} of seven O-type and eleven B-type stars resulted in the detection of two O-type and two B-type stars. These four stars were all detected at $10 \ \mathrm{GHz}$ while only one was detected at $2.3 \ \mathrm{GHz}$. The lack of detection at this higher frequency is attributed to thermal free-free absorption in the extended wind. In the context of our results, this free-free absorption would need to occur at radii not captured by our simulation or the winds of the observed stars would need to be much denser. As discussed above, our simulations suggest this is unlikely. Mass-loss rates and therefore wind densities of the stars observed may also be lower than that used in our simulation, resulting in lower fluxes. Another possible explanation for the lack of detection is that the magnetic confinement of the stellar wind has, in the manner of our synthetic radio results, reduced the spectral flux possibly below the sensitivity of the JVLA.

\cite{Kurapati2016} report that their theoretical, $\dot{M}_{\mathrm{th}}$, and observationally inferred, $\dot{M}_{\mathrm{ob}}$, mass-loss rates may vary by a factor of 3, as the $\dot{M}_{\mathrm{th}}$ are based on the models of \cite{Vink2000} which assume smooth spherical symmetry and no magnetic confinement. This is indeed the order of variability seen in our synthetic lightcurves and continuum emission, where we see deviation by a factor of 3 for the magnetic wind compared to the spherical wind.

Our results agree with the results of \cite{Kurapati2016} to within the uncertainty stated for their $\dot{M}_{\mathrm{th}}$. However, we draw attention to the fact that magnetic confinement of the wind introduces a dependency of the emission on not only the density profile but also the observing inclination and phase; as both spherical and cylindrical symmetry of the wind has been broken.




\section{Conclusions}

We have performed 3D isothermal MHD simulations of a magnetic rotating massive star with a non-zero dipole obliquity and predicted the synthetic radio/sub-mm observable lightcurves and continuum spectra for a frequency range compatible with ALMA. From these results we also compare the simulation mass-loss rate, to the inferred observed mass-loss rate calculated from the synthetic spectral flux.

Despite the lack of shock heating and cooling physics imposed by the isothermal assumption, spherical and cylindrical symmetry is broken due to the obliquity of the stellar magnetic dipole resulting in an inclination and phase dependence of both the synthetic spectral flux and corresponding inferred mass-loss rate. Both quantities vary by factors between 2 and 3 over a full rotational period of the star, illustrating the divergence from a symmetric wind.

We also show that radio emission with a constant spectral index agrees well with our numerical prediction for a spherical wind at $\nu~<~10^{3} \ \mathrm{GHz}$, however it is unable to capture the behaviour of emission at $\nu~>~10^{3} \ \mathrm{GHz}$. As such we caution the use of such constant spectral index models for predicting emission from non-spherical winds such as those which form around magnetic massive stars. Our results further show that sub-mm frequencies of $10^{2} \ \mathrm{GHz}$ and greater are required to probe the inner winds of massive magnetic stars. Such frequencies, at the sensitivity required for detection, are only available via the ALMA observatory at the present time.

We have also demonstrated that predicting mass-loss rates via observed radio/sub-mm emission is sensitively dependent not only on the observing inclination but also on the rotational phase. As the magnetic field fundamentally changes the stellar wind structure and the mass-loss rate calculation is also dependent on the density structure, any mass-loss rate prediction is a function of the wind magnetic confinement. This make the consideration of the stellar magnetic field vital for accurately assessing mass-loss rates from massive magnetic stars. The method for predicting $\dot{M}_{\ast}$, derived from the theory in WB75, is not directly capable of achieving this. Indeed we find that at an observing frequency of $900 \ \mathrm{GHz}$, $\dot{M}_{\mathrm{sim}}$ is over estimated by up to a factor of three. A situation further clouded by the addition of the phase and inclination dependence mentioned above.

As this is the first 3D MHD simulation of a massive star wind incorporating an oblique dipole, there is a large parameter and physical space left to study. Future work will extend the current model to adiabatic physics, allowing for shock heating and optically thin cooling of the gas. Both of these physical mechanisms are important for the generation of higher energy emission such as X-ray and H$_{\alpha}$.

\section{Acknowledgements}

The authors thank the reviewer for their helpful comments and suggestions; which improved the quality and content of the publication.

Personal thanks goes to Dylan Kee for many useful discussions on which improved the quality of the physics investigated in this study.

A.uD acknowledges support by NASA through Chandra Award number TM7-18001X issued by the Chandra X-ray Observatory Center which is operated by the Smithsonian Astrophysical Observatory for and on behalf of NASA under contract NAS8- 03060

The authors acknowledge support from the Science and Facilities Research Council (STFC).

Computations were performed using the University of Birmingham's BlueBEAR HPS service, which was purchased through HEFCE SRIF-3 funds. See http://www.bear.bham.ac.uk.


\bibliographystyle{mnras}
\bibliography{/Users/sdaley/Work/Reading/References/ExoplanetRefs/exoplanet_refs,/Users/sdaley/Work/Reading/References/MHDRefs/mhd_refs,/Users/sdaley/Work/Reading/References/CollidingWindsRefs/colliding_wind_refs,/Users/sdaley/Work/Reading/References/StellarWindRefs/stellar_wind_refs,/Users/sdaley/Work/Reading/References/SPMIRefs/SPMI_refs,/Users/sdaley/Work/Reading/References/OStarRefs/O_star_refs}

\label{lastpage}

\end{document}